%% file: master.tex
\documentclass[12pt]{article}
\usepackage{epsfig}
\usepackage{graphics}
\usepackage{a4}
\usepackage{color}

\begin{document}
\input{math_macros}

%---------------------------------------------------------
%
% Title:
%
\title{Calculation of Asymptotic and RMS\\
Kicks due to Higher Order Modes\\
in the 3.9~GHz Cavity}
\author{L.~Bellantoni (FNAL),\\
        H.~Edwards (FNAL/DESY),\\ 
        R.~Wanzenberg (DESY)}

\date{\today}
\maketitle

\setlength{\marginparsep}{-8cm}
\setlength{\marginparwidth}{8cm}
\marginpar{\vspace*{-10cm} Fermilab Note TM-2404-AD-APC\_TD
\\ Internal Report DESY M 08-01 Mar 2008}

\section*{Abstract}
\begin{center}
\begin{minipage}{12cm}
FLASH plans to use a ``third harmonic" ($3.9$~GHz) 
superconducting cavity to compensate nonlinear distortions
of the longitudinal phase space due to the sinusoidal curvature
of the the cavity voltage of the TESLA $1.3$~GHz cavities.
Higher order modes (HOMs) in the $3.9$~GHz have a significant
impact on the dynamics of the electron bunches in a long 
bunch train.  Kicks due to dipole modes can be 
enhanced along the bunch train depending on the frequency and
Q-value of the modes.  The enhancement factor for a constant 
beam offset with respect to the cavity has been calculated.  A
simple Monte Carlo model of these effects, allowing for 
scatter in HOM frequencies due to manufacturing variances, has
also been implemented and results for both FLASH and for an
XFEL-like configuration are presented.
\end{minipage}
\end{center}
\newpage
%
%---------------------------------------------------------
%\tableofcontents
%---------------------------------------------------------
\input introduction.tex

\input wakehoms.tex

\input effects.tex

\input toymc.tex

\input conclusion.tex

\clearpage 
\newpage
%\subsection*{Acknowledgment}
%We would like to thank 

\input references.tex

%\listoffigures
\end{document}

%% file: math_macros.tex
% R W's favorite macros
%
\newcommand{\dummyfig}[1]{ 
\begin{picture}(80,40)(0,0)
\put(0,0){   \framebox(80,40){ #1}   }
\end{picture}
}
%
%---------------------------------------------------------
\newcommand{\dmyfig}[3]{ 
\begin{center}
 \epsfig{file=Eps2/B19MEuntuned1.eps, width=4cm,angle=-90}
\end{center}
}
%
%---------------------------------------------------------
%
% definition of deflist environment
%
\newcommand{\deflabel}[1]{\bf #1\hfill}%
\newcommand{\kVnC}{
          \mbox{$\mbox{kV/nC}$}}
%---------------------------------------------------------
\newcommand{\Vec}[1]{
\mbox{ \boldmath $\hspace*{-1mm} #1 \hspace*{-1mm} $ } }
\newcommand{\Abs}[1]{ \left| #1 \right|}
\newcommand{\Real}[1]{{\rm Re}\left( #1 \right)}
\newcommand{\Imag}[1]{{\rm Im}\left( #1 \right)}
\newcommand{\trace}{ {\rm tr}\, } 
\newcommand{\FT} [1]{ \widetilde{#1}}
%
%---------------------------------------------------------
%
\newcommand{\beq}{\begin{equation}}
\newcommand{\eeq}{\end{equation}}
\newcommand{\bea}{\begin{eqnarray} }
\newcommand{\eea}{\end{eqnarray} }
%---------------------------------------------------------
%
\newenvironment{deflist}[1]%
{\begin{list}{}{\settowidth{\labelwidth}{\bf #1}%
\addtolength{\leftmargin}{\labelsep}%
\let\makelabel\deflabel}}%
{\end{list}}
%
%
% L B's favorite macros
%-----------------------------------------------------------------------
%     Publication things:
%
\def \ie    {\hbox{\it i.e.}}     
\def \etc   {\hbox{\it etc.}}
\def \ibid  {\hbox{\it ibid.}}
\def \vs    {\hbox{\it vs.}}
\def \eg    {\hbox{\it e.g.}}     
\def \cf    {\hbox{\it cf.}}
\def \etal  {\hbox{\it et al.}}
\def \via   {\hbox{\it via}}
\def \NIM   {Nucl. Instr. Meth.}
\def \PR    {Phys. Rev.}
\hyphenation{back-ground}
\hyphenation{brem-sstrah-lung}
\hyphenation{had-ron had-ronic}
\hyphenation{syn-chro-tron}
\hyphenation{system-atic}
%
% To make figure/table captions narrower than main text
\newlength{\capwidth}
\setlength{\capwidth}{\textwidth}
\addtolength{\capwidth}{-2.0cm}
\font\twelvebf=cmbx10 scaled\magstep 1
\font\twelverm=cmr10 scaled\magstep 1
\font\twelveit=cmti10 scaled\magstep 1
\font\tenbf=cmbx10
\font\tenrm=cmr10
\font\tenit=cmti10
\font\ninebf=cmbx9
\font\ninerm=cmr9
\font\nineit=cmti9
\font\eightbf=cmbx8
\font\eightrm=cmr8
\font\eightit=cmti8
\font\sevenrm=cmr7
%
%-----------------------------------------------------------------------
%     Math things:
%     \beq goes into math mode and numbers the equations; $ just goes 
%     into math mode and $$ goes into math mode and centers the 
%     equation.  \bbeq and \ebeq start and end eqnarray* mode which is
%     unnumbered equations in table format; use \bbeqn and ebeqn for
%     numbered equations in table format; use \nonumber\\ rather than
%     plain old \\ to squelch numbers on every line.
\def \beq   {\begin{equation}}
\def \eeq   {\end{equation}}
\def \bbeq  {\begin{eqnarray*}}
\def \ebeq  {\end{eqnarray*}}
\def \bbeqn  {\begin{eqnarray}}
\def \ebeqn  {\end{eqnarray}}
\def \Tr    {\mathop{\mathrm Tr}}
\def \Im    {\mathop{\mathrm Im}}
\def \Re    {\mathop{\mathrm Re}}
\def \vect  {\overrightarrow}
\def \twdl  {\widetilde}
\def \hat   {\widehat}
\def \partder#1#2  {\partial #1\over\partial #2}
\def \secder#1#2#3 {\partial^2 #1\over\partial #2 \partial #3}
%
%
%-----------------------------------------------------------------------
%     Common mathematical physics symbols
\def \omg#1 {\mbox {${\mathcal O}(#1)$}}
\def \avg#1 {$\left\langle #1\right\rangle$}
\def \to    {\rightarrow}
\def \bra#1 {$\left\langle #1\right|$}
\def \ket#1 {$\left| #1\right\rangle$}
\def \braket#1#2 {\left\langle #1\right. \left| #2\right\rangle}
\def \amp#1 {${\mathcalA}(#1)$}
\def \apgt  {\raisebox{-0.6ex}{$\stackrel{>}{\sim}$}}
\def \aplt  {\raisebox{-0.6ex}{$\stackrel{<}{\sim}$}}
% To do asymmetric errors: \pma{pos err}{neg err}
\def \pma#1#2 {\mbox{\raisebox{-0.6ex}
           {$\stackrel{\scriptstyle \;+\; #1}{\scriptstyle \;-\; #2}$}}}
%
%  To do continued decay chains, ie
%     \begin{tabbing}
%     xxxxxxxxxxxxxxxxxxxxxxx\= \kill
%     \>B$^- \longrightarrow  $ \= D$^{**0}$(2420)~ $\mu^- \nu$  \\
%     \>                        \> \dr \= \Dsp~\pim$^-$ \\
%     \>                        \>     \> \dr \= \Do~\pib$^+$ \\
%     \>                        \>     \>     \> \dr K$^- \pi^+$ \\
%     \end{tabbing}
\def \dr    {$\;^\mid\!\!\!\longrightarrow$}
%
%     Common particle physics symbols
\def \ev    {\,\mathrm eV}
\def \kev   {\,\mathrm keV}
\def \mev   {\,\mathrm MeV}
\def \gev   {\,\mathrm GeV}
\def \km    {\,\mathrm km}
\def \cm    {\,\mathrm cm}
\def \mm    {\,\mathrm mm}
\def \um    {\,\mu\mathrm m}
\def \ghz   {\,\mathrm GHz}
\def \mhz   {\,\mathrm MHz}
\def \khz   {\,\mathrm kHz}
\def \hz    {\,\mathrm Hz}
\def \picos {\,\mathrm ps}
\def \ns    {\,\mathrm ns}
\def \us    {\,\mu\mathrm s}
\def \ms    {\,\mathrm ms}
\def \pb    {\,\mathrm {pb}$^{-1}$}
\def \mrad  {\,\mathrm mrad}
\def \BR#1#2 {\mbox{Br}(#1$\to$#2)}
\def \JP     {\mathrm J$^{\mathrm P}$}
\def \dedx {d{\it E}/d{\it x}}
\def \rphi {$r$-$\phi$}
\def \Pt2  {${\mathrm P}_\perp^2$}

%% file: introduction.tex
% ############################################################
%
% FILE: introduction.tex
%
% ############################################################
%
\section{Introduction}
\label{sec-1}

FLASH plans to use a ``third harmonic" ($3.9$~GHz) superconducting cavity
to compensate nonlinear distortions of the longitudinal phase space due
to the sinusoidal curvature of the the cavity voltage of the TESLA 
$1.3$~GHz cavities.  Higher order modes (HOMs) in the $3.9$~GHz have a 
significant impact on the dynamics of the electron bunches in a long 
bunch train.  The analysis here seeks to determine what level of damping,
if any, is required.

In the case where the spacing of the bunches is near (but not exactly
the same as) a multiple of the period of a dipole HOM, kicks from the
HOM can be resonantly enhanced along the bunch train, depending on the
frequency and Q-value of the modes.  The enhancement factor for a constant 
beam offset with respect to the cavity can be expressed in a simple analytic
form, which also provides the energy loss from resonance effects with a
monopole mode.

A simple Monte Carlo model of these effects, allowing for scatter in HOM
frequencies due to manufacturing variances, has also been implemented and
results for both a FLASH-like and for an XFEL-like configuration are 
presented.

The beam parameters for our analyses are:

\begin{table}[h]
\begin{center}
\begin{tabular}{lcc}
Parameter                &    FLASH  &     XFEL-like injector  \\
\hline                
Bunch spacing            &  $1\mu s$ &  $200ns$   \\
Bunch charge             &     $1nC$ &  $1nC$     \\
Bunch length (1$\sigma$) &     $1ps$ &  $1ps$     \\
Beam energy              & $130\mev$ &  $500\mev$ \\
Bunch offset at entry    &     $1mm$ &  $1mm$     \\
Bunches per train        &    800    &   800      \\
Number of cavities       &      4    &    32      \\
\hline
\end{tabular}
\vspace{-0.1cm}
\caption{\label{beampars}
Beam parameters for typical FLASH and XFEL injector running.}
\end{center}
\end{table}

An important issue is the specification for how much kick can be tolerated
before lasing stops.  At FLASH, the spot size at the location where the
``third harmonic" cavities will be installed is about 0.2mm and the invariant
emittance's design value is 1 mm-mrad in both $x$ and $y$; in real operation
it is often twice that.  The beam divergence is thus about 20$\mu$rad in the
best case.  We use $\pm10\mu$rad as our target.  We do not have a target for
energy loss, but find small values for these effects in all cases.

Although we are primarily concerned with FLASH, our methods are completely
general and we have done some investigation of the situation for XFEL.
The optics for the XFEL are in a state of flux at this writing; our selection
of parameters here is perforce somewhat arbitrary.  The $\pm 10\mu$rad
requirement is not far from other parameter sets that are under consideration
at this writing.  A more detailed study of the XFEL requirements is being
undertaken by Yauhen Kot and Thorsten Limberg.  We do not here allow for
betatron phase advance between the cavities, and this effect will be larger at
the XFEL than at FLASH.

%% file: wakehoms.tex
% ############################################################
%
% FILE: wakehoms.tex
%
% ############################################################
%
\section{Wakefield due to HOMs}
\label{sec-2}
The purpose of this section is to define the parameters that are important
for the long range wake field calculation.  Our development follows 
reference \cite{Wan01} closely.

%----------------------------------------------------------------------------
%----------------------------------------------------------------------------
\subsection{Modes in a cavity}
\subsubsection{The electric and magnetic fields}
Consider a monopole ($m=0$) or dipole mode ($m=1$) mode with the 
frequency $f = \omega/(2 \, \pi) $ in a cavity with cylindrical symmetry.
One obtains in complex notation for the electric and magnetic field:
\bea
  \Vec{E}(r,\phi,z,t)  & = &   
\begin{array}[t]{ll}
  \left( \,\, \FT{ E_r^{(m)} }(r,z) \right.     \, \cos(m\,\phi) 
                           & \Vec{e_r} \\
       \vspace*{1mm}
      +  \,\,  \FT{ E_{\phi}^{(m)} }(r,z) \, \sin(m\,\phi) 
                           & \Vec{e_{\phi}}
                              \\
      +  \,\,   \FT{ E_z^{(m)} }(r,z)      \, \cos(m\,\phi) 
                           &  \Vec{e_z} 
                \left. \phantom{\FT{E_r^{(m)}}} \hspace*{-6mm} \right)
			                   \, \exp(-i \, \omega \, t)  \\
 \end{array}
\nonumber \\
          \\
 \Vec{B}(r,\phi,z,t) & = & 
\begin{array}[t]{ll}
  \left( \,\, \FT{ B_r^{(m)} }(r,z) \right.     \, \sin(m\,\phi) 
                           & \Vec{e_r}
                                        \\
      +  \,\,  \FT{ B_{\phi}^{(m)} }(r,z) \, \cos(m\,\phi) 
                           & \Vec{e_{\phi}}
                                         \\
      +  \,\,   \FT{ B_z^{(m)} }(r,z)      \, \sin(m\,\phi) 
                           &  \Vec{e_z}  
                 \left. \phantom{\FT{E_r^{(m)}}} \hspace*{-6mm} \right)
			       \, \exp(-i \, \omega \, t).   \\
 \end{array}
\nonumber
\eea

%---------------------------------------------------------------------------
\subsubsection{The loss parameter and $R/Q$}

The interaction of the beam with a cavity mode is characterized by the
loss parameter $k^{(m)}_{\|}(r)$ or by the quantity $R/Q$ \cite{Wakes}.
These parameters can be determinated from the numerically calculated fields
using the MAFIA post-processor \cite{MAFIA,CST}.
The longitudinal voltage for a given mode at a fixed radius $r$ is defined as
\beq
  V^{(m)}_{\|}(r) = \int_0^L \, dz \, \FT{ E_z^{(m)} }(r,z) 
                     \exp(-i \, \omega \, z/c),
\label{longV}
\eeq 
while the total stored energy is given by:
\beq
  U^{(m)} = \frac{\epsilon_0}{2} \,\,  \int d^3r \,\, \Abs{ \FT{ \Vec{E^{(m)}} } }^2.
\eeq 

From the voltage and stored energy the loss parameter and $R/Q$ 
can be calculated:
\bea
  k^{(m)}(r) & = & \frac{ \Abs{ V^{(m)}_{\|}(r)}^2 }{  4 \, U^{(m)}} 
 \nonumber \\
           \\
 \nonumber \\
  \frac{ R^{(m)}}{Q}  & = & \frac{1}{r^{2 \,m}} \frac{ 2 \,\,k^{(m)}(r) }{ \omega }.
 \nonumber
\eea
For monopole modes the superscript $(0)$ is usually omitted $R/Q = R^{(0)}/Q$.
Although our definitions include a radial dependence in the loss parameter, 
$R^{(m)}/Q$ is independent of the radius $r$ since it can be shown
(see \cite{Wakes,Wei83}) that $ V^{(m)}(r) \sim r^m$ and therefore
$k^{(m)}(r) \sim r^{2 \,m}$.

%---------------------------------------------------------------------------
\subsubsection{The geometry parameter $G_1$ and the Q-value}

The power $P_{sur}$ dissipated into the cavity wall due to the surface
resistivity $R_{sur}$ can be calculated from the tangential magnetic field:
\beq
   P_{sur} = \frac{1}{2} \,\, R_{sur}  \,\, \int dA \Abs{ H_{sur} }^2. 
\eeq
For a superconducting cavity the surface resistance is the sum
of the BCS (Bardeen, Cooper, Schrieffer) resistance $R_{BCS}$, which depends
on the frequency and the temperature, and a residual resistivity $R_0$.   
The BCS resistance $R_{BCS}$ scales with the square of the frequency $f$ and 
exponentially with the temperature $T$:
\beq
 R_{BCS}(f,T) \propto  \frac{f^2}{ T} \,\exp(-1.76 \, T_c/T).
\eeq
The less-well understood residual resistance $R_0$ adds directly to
$R_{BCS}$ but remains in the limit $T \to 0$.

The total damping of a cavity mode is not only determined by the surface 
losses but also by coupling to external waveguides (HOM-dampers).  Therefore
one has to distinguish the Q-value $Q_0$ which is defined above and the 
external Q-value $Q_{ext}$ which characterizes the coupling to external 
waveguides.  Typically, $Q_0 >> Q_{ext}$.

The geometry parameter $G_1$ \cite{wilson82} is defined as:
\beq
    G_1 = R_{sur}  \,\, Q_0.
\eeq
$G_1$ is a purely geometric quantity that is independent of the cavity material; it depends only on the mode and the
shape of the cavity that creates that mode.

%----------------------------------------------------------------------------
%----------------------------------------------------------------------------
\subsection{Wakefields}
\subsubsection{Wake potential}
%
%------------------------------------------------
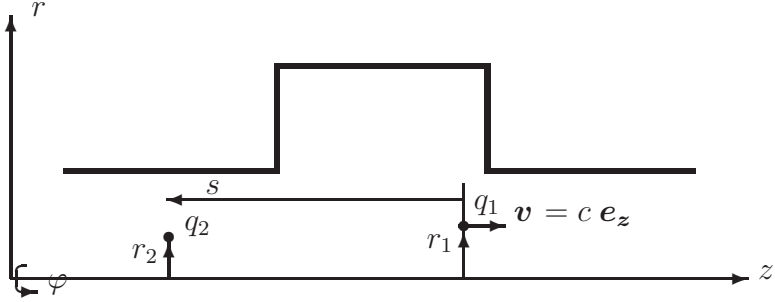
\begin{figure}[h!btp]
\setlength{\unitlength}{0.7mm}
\begin{center}
\begin{picture}(150,50)(0,10)
\thicklines
%\put(0,10){\dashbox{10}(150,80){ }}
%
\put(2,60){\makebox(0,0)[bl]{ $ r $}}
\put(0,10){\vector(0,1){50} }
\put(140,10){\makebox(0,0)[bl]{ $ z $}}
\put(0,10){\vector(1,0){140} }
\put(3,10){\oval(4,5)[l] }
\put(4,7.5){\vector(1,0){1} }
\put( 5,7.0){\makebox(0,0)[bl]{ $ \varphi $}}
%
% cavity
\put(50,50){\rule{41\unitlength}{1\unitlength}}
\multiput(50,30)(40,0){2}{\rule{1\unitlength}{20\unitlength} }
\multiput(10,30)(80,0){2}{\rule{40\unitlength}{1\unitlength} }
%
% bunch
\put(86,22){\makebox(0,0)[bl]{ $ q_1 $}}
\put(86,20){\circle*{2} }
\put(93,20){\makebox(0,0)[bl]{ $ \Vec{v} = c \, \Vec{e_z} $}}
\put(85,20){\vector(1,0){9} }
\put(86,10){\vector(0,1){9} }
\put(77,15){\makebox(0,0)[bl]{ $ r_1 $}}
\put(31,18){\makebox(0,0)[bl]{ $ q_2 $}}
\put(30,18){\circle*{2} }
\put(30,10){\vector(0,1){7} }
\put(21,13){\makebox(0,0)[bl]{ $ r_2 $}}
\put(86,25){\vector(-1,0){56.5} }
\put(86,20){\line(0,1){8} }
\put(35,26){\makebox(0,0)[bl]{ $ s $}}
\end{picture}
\end{center}

\caption {
A point charge $q_1$ traversing a cavity with an offset $r_1$ followed 
by a test charge $q_2$ with offset $r_2$ .
}
\label{pointwake}
\end{figure}
%------------------------------------------------

Consider the situation shown in Figure \ref{pointwake}. A test charge
$q_2$ follows a point charge $q_1$ at a distant $s$.
The distant $s$ is positive in the  direction opposite to the
motion of the point charge $q_1$.
Both charges are relativistic ($v \approx c$).
The Lorentz force on the test charge due to the fields generated
by the point charge $q_1$ is

\begin{equation}
   \Vec{F} = \frac{d\Vec{p}}{dt} = \,\,
         q_2 \, ( \Vec{E} + c \, \Vec{e_z} \times \Vec{B} ).
\end{equation}
The {\it wake potential } of the point charge $ q_1 $ is defined as:
\begin{equation}
     \Vec{W}(x_2,y_2,x_1,y_1,s) = \frac{1}{q_1} \,\,
        \int_0^L dz ( \Vec{E}
          + c \, \Vec{e_z} \times \Vec{B} )_{t = (z+s)/c}.
\end{equation}
The wake potential is the integrated Lorentz force
on a test charge. Causality requires $\Vec{W}(s) = 0$ for $s<0$.

The longitudinal and transverse components of the 
wake potential are connected by the Panofsky-Wenzel theorem \cite{PaWe56}
\begin{equation}
\frac{\partial}{\partial s} \Vec{W}_{\bot}(x_2,y_2,x_1,y_1,s)
= - \Vec{\nabla}_{\bot_2} W_{\|} (x_2,y_2,x_1,y_1,s).
\end{equation}
Integration of the transverse gradient (applied to the transverse
coordinates of the test charge) of the longitudinal
wake potential yields the transverse wake potential.

%--------------------------------------------------------------
\subsubsection{Multipole expansion of the wake potential}
If the structure traversed by the bunch is cylindrically 
symmetric then a multipole expansion can be used to describe the
wake potential.  The location of the bunch train in the $(r,\phi)$ plane
will break the symmetry and determine the azimuthal orientation
of the $m > 0$ modes.
Consider again the situation shown in Figure \ref{pointwake}. 
Assume that the point charge $q_1$ traverses the cavity
at position $(r_1,\varphi_1)$, while the test charge follows at 
position $(r_2,\varphi_2)$.
The longitudinal wake potential may be expanded in multipoles:
\begin{equation}
W_{\|}(r_1,r_2,\varphi_1,\varphi_2,s) =
\sum_{m=0}^{\infty} \, {r_1}^m \, {r_2}^m \,\,
 W^{(m)}_{\|}(s)
 \,\, \cos m\,( \varphi_2 - \varphi_1) .
\label{wakecyl}
\end{equation}
The functions 
$W^{(m)}_{\|}(s)$ are the longitudinal $m$-pole wake potentials.
There is no a-priori relation between the wake potentials
of different azimuthal order $m$.  

The transverse wake potential can be calculated using the
Panofsky-Wenzel theorem, and the transverse $m$-pole wake 
potentials are defined as:
\beq
  W^{(m)}_{\bot}(s)  =  -  \int_{-\infty}^{s}ds' \, W^{(m)}_{\|}(s'),
\label{Wmtrans}
\eeq
for $m > 0$. There is no transverse monopole wake potential.
The dipole wake potential does not depend on the position of the
test charge $q_2$. The kick on the test charge is linear in
the offset of the point charge $q_1$.

%----------------------------------------------------------------------------
\subsubsection{Wakefields due to HOMs}

It is possible to write the m-pole
wake potentials $W^{(m)}_{\|}(s)$ as a sum over all modes:
\bea
\label{wake1}
  W^{(m)}_{\|}(s) & = & - \sum_n \, \, \omega_n
                    { \left( \frac{R^{(m)}}{Q} \right) }_n
                    \cos( \omega_n \, s/c) \,\,
                    \exp( - 1/{\tau_n} \,\,s/c) 
                    \nonumber \\
  & & \\
  W^{(m)}_{\bot}(s) & = & c \, \sum_n \, \,
                    { \left( \frac{R^{(m)}}{Q} \right) }_n
                    \sin( \omega_n \, s/c) \,\,
                    \exp( - 1/{\tau_n} \,\,s/c). 
                    \nonumber 
\eea
where $ \omega_n $ are the frequencies of the m-pole modes.
A damping term has been included with the damping time $\tau_n$ for mode $n$.
As  $Q_0 >> Q_{ext}$, the damping time of the voltage is very nearly
\beq
             \tau_n \approx \frac{2 \, (Q_{ext})_n}{\omega_n}.
\eeq

%% file: effects.tex
% ############################################################
%
% FILE: effects.tex
%
% ############################################################
%
\section{Effects of long range wakefields on a bunch train}
\label{sec-3}
%
%--------------------------------------------------------------------
%--------------------------------------------------------------------
% 

\subsection{Energy deviations and kicks on the bunches}
\begin{figure}[h!tbp]
\setlength{\unitlength}{1mm}
%---------------------------------------
\centering
%\dummyfig{./Eps/btrain.eps}
\includegraphics*[width=120mm]{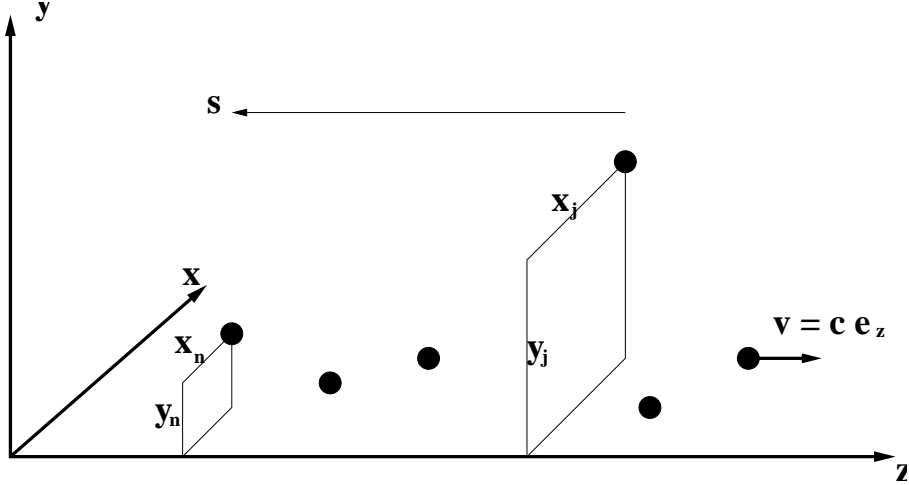}
%---------------------------------------
%\vspace*{-4mm}
\caption{Schematic representation of a train of bunches with
         offsets $x_i$ and $y_i$ with respect to the reference axis
         of the accelerator.}
\label{btrain}
\end{figure}

The long range wakes due to HOMs can cause energy deviations
and kicks on the bunches. A bunch train of $N$ bunches is 
shown in Figure \ref{btrain}; the notation for the offsets with
respect to the reference axis of the accelerator is
$x_i$ and $y_i$ and the direction of the longitudinal
coordinate $s$ ($s=0$ at the first bunch of the train) is also shown.
It is assumed that all bunches have the same bunch charge $q$.
In our investigation of the energy deviation and the kick on the bunch
$n$ within the bunch train we profited strongly from the analysis
of long range wakes and beam loading by P.~Wilson \cite{wilson82}.
%-------------------------------------------------
%-------------------------------------------------

The energy deviation of bunch $n$ due to HOMs is a sum over all
HOMs produced by all preceding bunches in the train. Additionally the
the self-wake due to HOMs is also included ($W^{(0)}_{\|}(0)$ and
$W^{(1)}_{\|}(0)$), which is the HOM equivalent of the fundamental
theorem of beam loading \cite{wilson82}.
\bea
\Delta E(s_n) & = & e \, q \, W^{(0)}_{\|}(0) \\
              &   & + e \, q \sum_{j<n} \,  W^{(0)}_{\|}(s_n-s_j)
	                                                \nonumber  \\
              &   &     \nonumber \\
	      &   & + e \, q \, ({x_n}^2 + {y_n}^2) W^{(1)}_{\|}(0)
	                                              \nonumber \\					      
              &   & + e \, q \sum_{j<n} \,\, 
                           (x_n \,  x_j + y_n \,  y_j) \,\,
                           W^{(1)}_{\|}(s_n-s_j)
                           \nonumber \\
              &   & + \dots  \nonumber 
\eea

As an example the energy deviation due the selfwake of a higher monopole
mode with a frequency of $f = 6$~GHz is considered. The contribution to
the energy deviation for a bunch with a
population of $N = 6.5 \times 10^9$ electrons is given by:
\beq
  \Delta E = e \, 1 \,{\rm nC} \,\, \, 2 \,\pi \,\, 6 \, {\rm  GHz} \,\, 
                   \frac{1}{2} \,\,  \frac{R}{Q}.
\eeq
An energy deviation of $196$~eV corresponds to an impedance ($R/Q$) of
10~$\Omega$ and a loss parameter of $k^{(0)}_{\|} = 0.188$~V/pC.
The factor $1/2$ is a consequence of the fundamental theorem of beam loading.

Now consider the transverse long range wakefields.  The kick on bunch $n$ due to
the dipole wake field is:
\bea
  \Vec{\theta}_n & = &\frac{e \, q_{bunch}}{E_{beam}} \,\,
                       \sum_{j<n} \,\, 
                       ( x_j \,\, \Vec{e_x} + y_j \,\, \Vec{e_y} )
                       W^{(1)}_{\bot}(s_n-s_j) \\
                &   & + \dots  \nonumber,
\eea
where $E_n$ is the energy of bunch.
The kick on bunch $n$ due to {\bf one} dipole mode with
revolution frequency $\omega_1$ and damping constant $\tau_1$ is:
\beq
  \Vec{\theta}_n =  \hat{\theta}_n \,\,
                       \sum_{j<n} \,\, 
                       ( \frac{x_j}{x_0} \,\, \Vec{e_x}
		        + \frac{y_j}{x_0} \,\, \Vec{e_y} )
		         \sin( \omega_1 \, (s_n-s_j)/c) \,\,
                    \exp( - 1/{\tau_1} \,\,(s_n-s_j)/c), 
\eeq
where the kick amplitude $\hat{\theta}_n$ on bunch $n$ is defined as
\beq
\hat{\theta}_n = \frac{e \, q_{bunch}}{E_{beam}} \, c \, \frac{R^{(1)}}{Q} \, r_0,
\label{Eqn_theta}
\eeq
with respect to an arbitrary reference offset $r_0$.

%--------------------------------------------------------------------
%--------------------------------------------------------------------
\subsection{One dipole mode and a bunch train with constant offset}
\label{sec-31}
It is instructive to consider the simplified situation of a bunch
with a constant offset with respect to the ``third harmonic" cavity. This
corresponds to an injection error or an misalignment of the cavity.
Since the long range dipole wakefield is a linear superposition of
HOMs it is sufficient to consider only one mode at a time in all
analytical formulas. 

Modes from the first 3 dipole passbands with the highest values for
$R^{(1)}/Q$ are summarized in table \ref{dipolmodes}, taken from reference
\cite{TimerCalculates}, 
along with high $R^{(m)}/Q$ modes of other azimuthal number.
The kick amplitude $\hat{\theta}$ for the dipole modes has been calculated
according to Equation \ref{Eqn_theta} assuming that all bunches have the same
energy of 130~MeV, bunch charge of 1~nC and reference offset of 1~mm.

%%%%%%%%%%%%%%%%%%%%%%%%%%%%%%%%%%%%%%%%%%%%%%%%%%%%%%%%%%%%%%%%%%%%%%%%%%
\begin{table}[h!tb]
\centerline{
 \begin{tabular} {c|c|c|c|c|c}
$f$ / GHz & $m$ & $R^{(m)}/Q$ /            & $G_1$ /  & $k^{(m)} / r_0^{(2m)}$ /               & $\hat{\theta}$ / $\mu$rad\\
          &     & $\Omega/{\rm cm}^{(2m)}$ & $\Omega$ & ${\rm V}/({\rm pC} \,{\rm cm}^{(2m)})$ & \\ 
 \hline
%   f / GHz  m  R/Q /Ohm/cm2  G1/Ohm  k1    theta
    7.506 &  0 &    23.3 &  475.5  & 0.55  &      \\
    4.834 &  1 &    50.7 &  277.3  & 0.77  & 1.22 \\
    5.443 &  1 &    20.9 &  426.2  & 0.36  & 0.50 \\
    7.669 &  1 &    29.5 &  470.7  & 0.71  & 0.71 \\
    9.133 &  2 &    11.2 &  402.9  & 0.32  &      \\
\end{tabular}
}
\vspace{-0.1cm}
\caption{\label{dipolmodes} 
RF-parameters and kick amplitude of modes with high $R^{(m)}/Q$.  For the dipole
modes ($m = 1$), the kick amplitude has been calculated for an beam energy of 130~MeV,
a bunch charge of 1~nC and an reference offset ($r_0$) of 1~mm.
}
\label{Tbl_modes}
\end{table}
%%%%%%%%%%%%%%%%%%%%%%%%%%%%%%%%%%%%%%%%%%%%%%%%%%%%%%%%%%%%%%%%%%%%%%%%%%
Furthermore it is now assumed that the bunch to bunch distance $\Delta t$ is
constant:
\beq
\Delta t = \frac{\Delta s}{c} = n_{fb} \, \frac{1}{f_{fu}},
\eeq
where $f_{fu} = 3.9$~GHz is the frequency of the fundamental mode and
$n_{fb}$ is the number of free buckets between bunches. The following
bunch distances have to be considered for the operation of the injector linear
accelerator:
%%%%%%%%%%%%%%%%%%%%%%%%%%%%%%%%%%%%%%%%%%%%%%%%%%%%%%%%%%%%%%%%%%%%%%%%%%
\begin{table}[h!tb]
\centerline{
 \begin{tabular} {r|c|r}
 $\Delta t$ / ns & $1/\Delta t$ / MHz & $n_{fb}$ \\
\hline
   200 & 5.0 &   780 \\
  1000 & 1.0 &  3900 \\
  2000 & 0.5 &  7800 \\
 10000 & 0.1 & 39000 \\
\end{tabular}
}
\vspace{-0.1cm}
\caption{\label{deltat} 
Typical bunch to bunch distance for the operation of injector linear
accelerator
}
\end{table}
%%%%%%%%%%%%%%%%%%%%%%%%%%%%%%%%%%%%%%%%%%%%%%%%%%%%%%%%%%%%%%%%%%%%%%%%%%

The bunch distance can be translated into a phase distance $\delta$ between
bunches:
\beq
 \delta = \omega_1 \, \Delta t = 2 \pi \frac{f_1}{f_{fu}} \, n_{fb},
\eeq
where $\omega_1 =  2 \pi \, f_1$ is the frequency of the considered
dipole mode.

A small change in the dipole frequency
due to fabrication tolerances will cause a large change in the bunch
to bunch phase since the number of free buckets is relatively
large. One obtains for a bunch distance of $1/\Delta t = 1 {\rm MHz}$:
\bea
 \Delta \delta & = & 2 \pi \frac{1}{f_{fu}} \, n_{fb} \, \,\Delta f_1 \\
               &   & \nonumber \\
	       & = & 0.36^{\circ} \, \frac{\Delta f_1}{{\rm kHz}}
\eea
A change of $10^{\circ}$ in the bunch to bunch phase after $n_{fb} = 3900$ 
free buckets corresponds to a frequency shift of about 20~kHz, which is much
smaller than the expected variation due to manufacturing variations.

\begin{figure}[h!tbp]
\setlength{\unitlength}{1mm}
%---------------------------------------
\centering
%\dummyfig{./Eps/dversusQ.eps}
\includegraphics*[width=120mm]{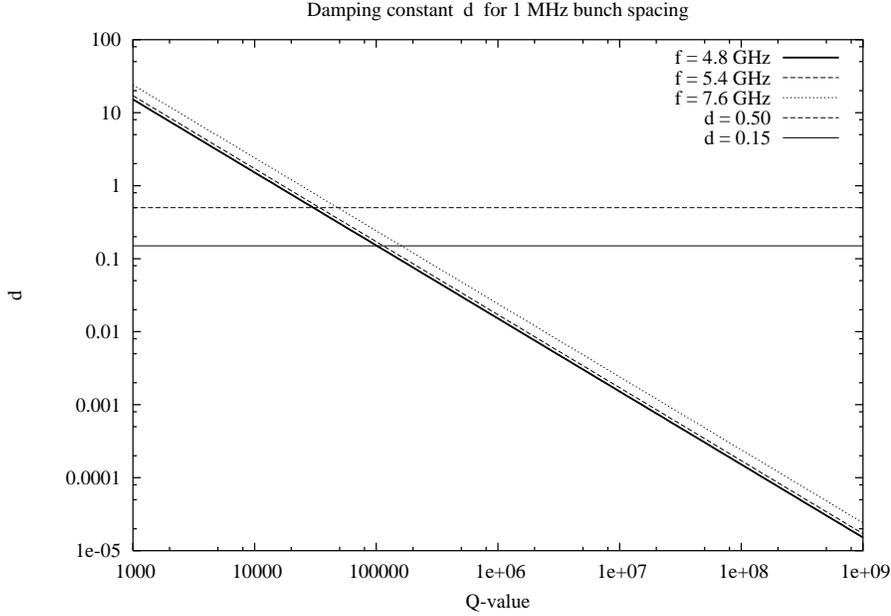}
%---------------------------------------
%\vspace*{-4mm}
\caption{The damping constant d versus the Q-value
for the three dipole modes from table \ref{dipolmodes}.
Additionally the damping constants $d=0.5$ and $d=0.15$ are
indicated with parallel lines.}
\label{dvq_plot}
\end{figure}

A bunch to bunch damping constant for the kick voltage is defined as:
\beq
 d = \frac{\omega_1}{2 \, Q_1} \, \Delta t
   = 2 \pi \frac{f_1}{f_{fu}} \, n_{fb} \, \frac{1}{2 \, Q_1},
\eeq
where $Q_1$ is the Q-value of the dipole mode, which is usually
dominated by the external Q-value of the HOM damper.
We give most of our expressions both in terms of $d$ and of $a = e^{-d}$.
A plot of the damping constant as a function of the Q-value is
shown in Figure~\ref{dvq_plot} in a double logarithmic scale for
the three modes considered in table \ref{dipolmodes} and
a bunch to bunch spacing of $1/\Delta t =1$~MHz.
%-------------------------------------------------
%-------------------------------------------------
With HOM dampers
Q-values of about $10^5$ are achieved, which corresponds to a damping
constant of about $0.15$ ($1/\Delta t =1$~MHz). If no HOM-dampers are
mounted on the cavity the Q-value will be larger than $10^9$ and the
damping constant $d$ will be very small ($ \sim 10^{-5}$).
Some details of the damping constants for the modes from
Table \ref{dipolmodes} are listed in the Table \ref{dipolmodesdamping}.
%%%%%%%%%%%%%%%%%%%%%%%%%%%%%%%%%%%%%%%%%%%%%%%%%%%%%%%%%%%%%%%%%%%%%%%%%%
\begin{table}[h!tb]
\centerline{
 \begin{tabular} {c||c|c|c|c|c|c|c}
 & \multicolumn{3}{c|}{$1/\Delta t =1$~MHz} & &
 \multicolumn{3}{c}{$1/\Delta t =5$~MHz}\\
 $f$ / GHz & \multicolumn{3}{c|}{Q-Value} & &\multicolumn{3}{c}{Q-Value}\\
  & $10^4$ & $5 \times 10^4$ & $10^5$ & & $10^4$ & $5 \times 10^4$ & $10^5$\\  
 \hline
4.834 &  1.5 & 0.30 & 0.15 &  &  0.30 & 0.06 & 0.04 \\
5.443 &  1.7 & 0.34 & 0.17 &  &  0.34 & 0.07 & 0.03 \\
7.669 &  2.4 & 0.48 & 0.24 &  &  0.48 & 0.10 & 0.05 \\
\end{tabular}
}
\vspace{-0.1cm}
\caption{\label{dipolmodesdamping} 
Damping constants $d$ for three modes and different external Q-value
and bunch-to-bunch distances of $1000$~ns and $200$~ns.
}
\end{table}
%%%%%%%%%%%%%%%%%%%%%%%%%%%%%%%%%%%%%%%%%%%%%%%%%%%%%%%%%%%%%%%%%%%%%%%%%%
Using the above defined parameters $\delta$ and $d$, the energy deviation
and the kick on bunch $n \in \{1,2,3\ldots\}$ due to one dipole mode are:
\bea
\Delta E_n & = &  \hat{E} \left(
                  \frac{1}{2} + \sum_{j=1}^{n-1}
		  \cos(\delta \, (n-j)) \,
		  \exp(- d \, (n-j)) 
		  \right) \label{En_mode}, \\
		  \nonumber \\
\theta_n & = & \hat{\theta} \,\,\, \sum_{j=1}^{n-1}
		   \sin(\delta \, (n-j)) \,
		   \exp(- d \, (n-j)), 
		   \label{theta_mode}		  
\eea

where
\beq
\hat{E} = - e \, q \, \omega_1 \frac{R^{(1)}}{Q} \, {r_0}^2,
\hspace*{5mm} {\rm and} \hspace*{5mm}
\hat{\theta} = \frac{e \, q}{E_0/c} \, \frac{R^{(1)}}{Q} \, r_0.
\eeq
\noindent $r_0$ is a reference offset.
The above expression (\ref{En_mode}) and (\ref{theta_mode}) can be rewritten as
\bea
\Delta E_n & = &  \hat{E} \left(
                  \frac{1}{2} + \Real{S_n} \right), \nonumber\\
		  \nonumber \\
\theta_n & = & \hat{\theta} \,\,\, \Imag{S_n},		  
\eea
with a sequence of complex sums $S_n$, defined as
\beq
   S_n = \sum_{j=1}^{n-1} \exp\left( (n-j) \, D  \right)
       = \sum_{j=1}^{n-1} \exp\left( j \, D  \right). 
\eeq
The complex damping constant $D$ is defined as $D = i \, \delta - d $.
The sequence $S_n$ may be calculated via a recurrence relation:
\bea
 S_1 & = & 0 \\
 S_{n+1} & = & \left( S_{n} + 1 \right) \exp(D), \nonumber
\eea
or via an explicit expression for the sum of a geometric series:
\beq
S_{n}  =  \frac{ 1 - \exp((n-1) \,D) }{\exp(-D) -1} 
\longrightarrow  \frac{ 1}{\exp(-D) -1}, 
\,\, {\rm for} \, \, n \rightarrow \infty.
\label{Eqn_Sn}
\eeq
Furthermore let $R_n$ be defined as:
\beq
R_{n}  =  \frac{\exp((n-1) \,D) }{\exp(-D) -1}
\label{Rndef}
\eeq
so that $$S_n = \lim_{n \rightarrow \infty} S_n - R_n.$$
The explicit expressions $\Real{R_n}$ and $\Imag{R_n}$ are
\bea
\Real{R_n} & = & \frac{ {\rm e}^{- n\,d} \left( \cos(n \, \delta) -
                       {\rm e}^{-d} \, \cos((n-1) \, \delta) \right)}{
                   1 - 2 \, {\rm e}^{-d} \cos(\delta) + {\rm e}^{- 2\,d} } \\
                \nonumber \\
\Imag{R_n} & = & \frac{ {\rm e}^{- n\,d} \left( \sin(n \, \delta) -
                       {\rm e}^{-d} \, \sin((n-1) \, \delta) \right)}{
                   1 - 2 \, {\rm e}^{-d} \cos(\delta) + {\rm e}^{- 2\,d} }.
\eea

%
%--------------------------------------------------------------------
%--------------------------------------------------------------------
\subsection{The functions $F_{R,n}$ and $F_{I,n}$}
\label{sec-3n}
\begin{figure}[h!tbp]
\setlength{\unitlength}{1mm}
%---------------------------------------
\centering
%\dummyfig{./Eps/d015FRFI.eps}
\includegraphics*[width=120mm]{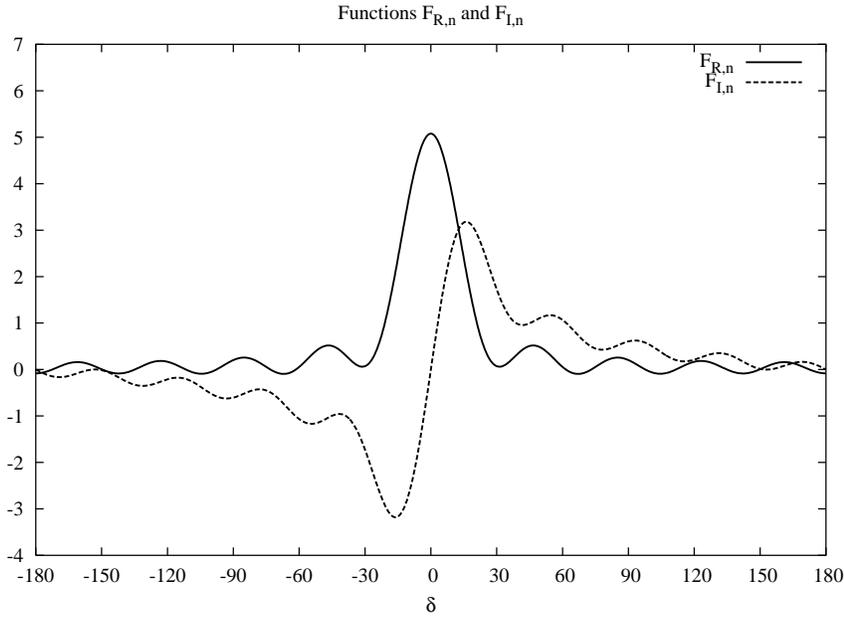}
%---------------------------------------
%\vspace*{-4mm}
\caption{The functions $F_{R,n}(\delta,d)$ and $F_{I,n}(\delta,d)$
versus the bunch to bunch phase $\delta$ for
$d=0.15$ and $n=10$ bunches.}
\label{d015FRNFIN_plot}
\end{figure}

The energy deviation $\Delta E_n$ of bunch $n$ and the kick $\theta_n$
on bunch number $n$ caused by the previous bunches are
$$\Delta E_n = \hat{E}(\frac{1}{2} - \Real{S_n}) = \hat{E} F_{R,n}(\delta,d)$$
and
$$\theta_n = \hat{\theta} \, \Imag{S_n} = \hat{\theta} F_{I,n}(\delta,d).$$
These equations define the functions $F_{R,n}(\delta,d)$ and
$F_{I,n}(\delta,d)$, which depend on the bunch to bunch phase
advance $\delta$ and the damping constant $d$.
A plot of the functions $F_{R,n}(\delta,d)$ and $F_{I,n}(\delta,d)$
is shown in Figure~\ref{d015FRNFIN_plot} for bunch number 10 and
for a damping constant $d=0.15$.
%-------------------------------------------------
%-------------------------------------------------

We have the following explicit expressions for these functions:
\bea
F_{R,n}(\delta,d) & = &
\frac{{\frac{(1 - a^2 )}{2} -a^n\cos(n\delta) +a^{n+1}\cos((n-1)\delta }}
	     {(1 - 2a\cos(\delta)  + a^2 )} \\
& &   \nonumber \\
F_{I,n}(\delta,d) & = &
	\frac{{a\sin(\delta) -a^n\sin(n\delta) +a^{n + 1}\sin((n-1)\delta }}
	{(1 - 2a\cos(\delta)  + a^2 )}.
\eea

The bunch to bunch phase can be regarded as a random quantity since the 
interbunch phase difference $\delta$ variation from expected variations
in manufacturing is much larger than $\pm 2 \pi$.  Consequently, it is
valuable to calculate the average and the rms value
of $F_{R,n}(\delta,d)$ and $F_{I,n}(\delta,d)$ over the range 
$-\pi \leq \delta \leq \pi$. We are also interested in the absolute
kick amplitude which is measured as the average of $\Abs{F_{I,n}(\delta,d)}$.
These are:
\bea
 \left<F_{R,n} \right> & = & \frac{1}{2 \, \pi}\int_{-\pi}^{\pi} d\delta \, F_{R,n}(\delta,d)
                             =  \frac{1}{2} \\
		   \nonumber \\
 \left<F_{I,n} \right> & = & \frac{1}{2 \, \pi} \int_{-\pi}^{\pi} d\delta \, F_{I,n}(\delta,d)
                             =  0 \\
		   \nonumber \\
 \left<\Abs{F_{I,n}} \right>
                   & = & \frac{1}{\pi} \int_{0}^{\pi} d\delta \, F_{I,n}(\delta,d) \nonumber \\
                   & = & \frac{1}{\pi} \sum_{k=1}^{n-1} \frac{1-(-1)^k}{k} \exp(-k \, d) \nonumber \\
                   & = & \frac{1}{\pi} \left( \frac{\exp(-n \, d)}{n} H(n,d) + \ln(\coth(\frac{d}{2})) \right),	  
\eea
where the function $H(n,d)$ is defined in terms of the hypergeometric function $_{2}F_{1}$
\cite{Abramowitz} as:
\beq
H(n,d) = {(-1)}^n \, {_{2}F_{1}}(1,n;n+1;-\exp(-d)) - \, {_{2}F_{1}}(1,n;n+1;\exp(-d))
\eeq
                      
Furthermore one obtains for the RMS-values
\bea
{\rm RMS} (F_{R,n}) & = &
	\sqrt{\frac{1}{2\,\pi} \int_{-\pi}^{\pi} d\delta \, F_{R,n}(\delta,d)^2 }
		\nonumber \\
		    & = &
	\sqrt{\frac{1}{4} + \frac{\exp(-n d)}{2} \frac{ \sinh((n-1)d) }{ \sinh(d) }  }
		\nonumber \\
                    & = &
         \sqrt{\frac{1+a^2-2 \, a^{2n}}{4 \, (1-a^2)}}
                          \\
{\rm RMS} (F_{I,n}) & = &
	\sqrt{\frac{1}{2\,\pi} \int_{-\pi}^{\pi} d\delta \, F_{I,n}(\delta,d)^2 }
		\nonumber \\
                    & = &
	\sqrt{\frac{\exp(-n d)}{2} \frac{ \sinh((n-1)d) }{ \sinh(d) }  }
\label{Eqn_FinRMS}
                          \\
                    & = &
	\sqrt{ \frac{1}{2} \frac{a^2-a^{2n}}{(1-a^2)}  }.
                 \nonumber
\eea

For small $(\ll n)$ values of $d$ we have:
\beq
{\rm RMS} (F_{I,n})
%\approx \sqrt{\frac{n-1}{2} (1 - n \, d)}
\approx  \sqrt{\frac{n-1}{2}} \;\;\; {\rm for} \, \, d \rightarrow 0.
\eeq

\begin{figure}[h!tbp]
\setlength{\unitlength}{1mm}
%---------------------------------------
\centering
%\dummyfig{./Eps/FRFI.eps}
\includegraphics*[width=120mm]{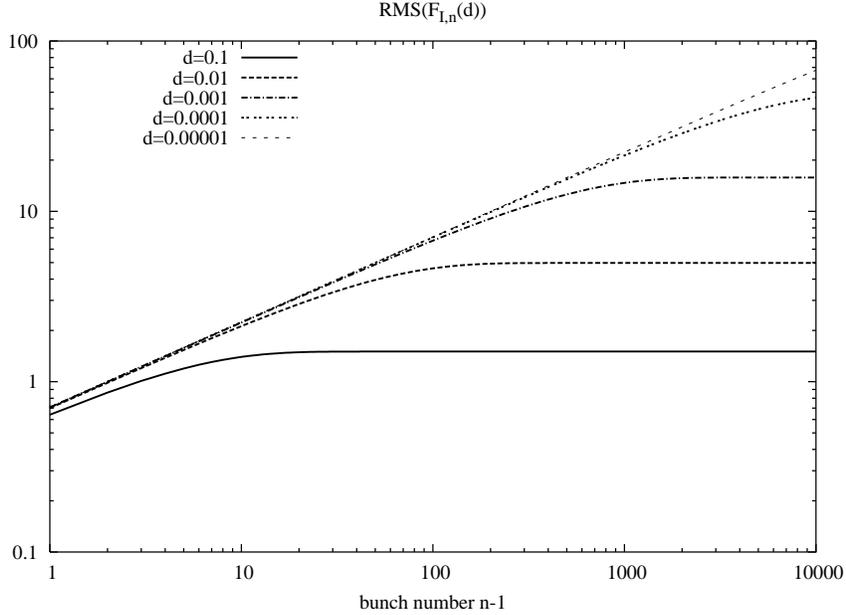}
%---------------------------------------
%\vspace*{-4mm}
\caption{The functions ${\rm RMS} (F_{I,n})$ 
versus $n-1$ for different damping constants $d$.}
\label{RMSFIN_plot}
\end{figure}

Because of the appearance of $n-1$ in the square root here, it is instructive
to plot ${\rm RMS} (F_{I,n})$ as function of the bunch number minus 1.
A plot of ${\rm RMS} (F_{I,n})$ for five different damping constants $d$
in the range from $0.1$ to $10^{-5}$ is show in Fig.~\ref{RMSFIN_plot}
for bunch numbers $n-1$ from 1 to $10^5$.
%-------------------------------------------------
%-------------------------------------------------
%\clearpage
%\newpage
%--------------------------------------------------------------------
The RMS-value of $F_{I,n}$ as a function of $n$ quickly
approaches the RMS-value
of the asymptotic function $F_I$ if the damping constant is relatively
large. To be more precise we define the ratio $r$ of the RMS-values as:
\beq
\frac{ {\rm RMS}(F_{I,n}) }{\lim_{n\rightarrow \infty} {\rm RMS} (F_{I,n})} = 
\sqrt{1 - a^{2 \, (n-1)} } = r < 1.
\label{RMSratio}
\eeq
Equation (\ref{RMSratio}) can be solved for the bunch $n$:
\beq
n = \left\lceil 1 + \frac{-\ln(1-r^2)}{2 \, d} \right\rceil,
\label{n_versus_d}
\eeq
where the bracket $\lceil\;\rceil$ indicates the smallest integer not less than
the expression in the bracket (ceiling function).
In Tab.~\ref{bunch_r095} we have summarized the bunch numbers $n$
for which ${\rm RMS} (F_{I,n})/{\rm RMS} (F_{I})$ is larger or equal
$r = 0.95$ using the damping constants from Fig.~\ref{RMSFIN_plot}.
%%%%%%%%%%%%%%%%%%%%%%%%%%%%%%%%%%%%%%%%%%%%%%%%%%%%%%%%%%%%%%%%%%%%%%%%%%
\begin{table}[h!tb]
\centerline{
 \begin{tabular} {c||c|c|c|c|c}  
d & 0.1 & 0.01 & 0.001 & 0.0001 &  0.00001 \\
 \hline
n & 13  & 118  & 1165  &  11641 &  116397  \\
\end{tabular}
}
\vspace{-0.1cm}
\caption{\label{bunch_r095} 
Bunch number $n$ for which the ratio of RMS values of the function
$F_{I,n}$ and $F_I$ is equal or larger than $r=0.95$.
}
\end{table}
%%%%%%%%%%%%%%%%%%%%%%%%%%%%%%%%%%%%%%%%%%%%%%%%%%%%%%%%%%%%%%%%%%%%%%%%%%
For a given damping constant $d$ we can find the bunch number $n$
according to Eqn. (\ref{n_versus_d}) for which the RMS-value of asymptotic function
$F_I$ is a good approximation to the RMS-value of the function $F_{I,n}$.
For a bunch train with 800 bunches the asymptotic function is only
a useful approximation if the damping constant is larger or equal to
$0.01$ which corresponds to a Q-value which is smaller or equal than $10^6$. 

%--------------------------------------------------------------------
\subsection{The asymptotic functions $F_R$ and $F_I$}
\label{sec-32}
For nearly all cases of interest, the functions $F_{R,n}(\delta,d)$ and
$F_{I,n}(\delta,d)$ reach their large-$n$ asymptotic limits 
$F_R(\delta,d)$ and $F_I(\delta,d)$ after only a small fraction of the 
bunch train has gone through the cavities.   The expressions in this section
are then useful.
\bea
F_R(\delta,d) & =  & \frac{1}{2} + \lim_{n \rightarrow \infty} \Real{S_n} 
                 \nonumber \\
              & =  & \frac{1 - {\rm e}^{- 2\,d} }{
                     2 \, ( 1 - 2 \, {\rm e}^{-d} \cos(\delta) + {\rm e}^{- 2\,d}) } 
                 \nonumber \\
              & =  & \frac{\sinh(d)}{
                     2 \, ( \cosh(d) - \cos(\delta) ) } 
                 \nonumber \\
              & =  & \frac{1 - a^2}{
                     2 \, (1 - 2 a\cos(\delta) + a^2) } 
		     \label{FRdef} \\
& &   \nonumber \\
F_I(\delta,d) & =  & \lim_{n \rightarrow \infty} \Imag{S_n} 
                 \nonumber \\
              & =  & \frac{{\rm e}^{-d} \sin(\delta) }{
                  1 - 2 \, {\rm e}^{-d} \cos(\delta) + {\rm e}^{- 2\,d} }
                 \nonumber \\
              & =  & \frac{ \sin(\delta) }{
                  2 \, (\cosh(d) - \cos(\delta)) }.
                 \nonumber \\
              & =  & \frac{a \sin(\delta)}{
                     (1 - 2 a\cos(\delta) + a^2) } 
		     \label{FIdef}
\eea

\begin{figure}[h!tbp]
\setlength{\unitlength}{1mm}
%---------------------------------------
\centering
%\dummyfig{./Eps/d015FRFI.eps}
\includegraphics*[width=120mm]{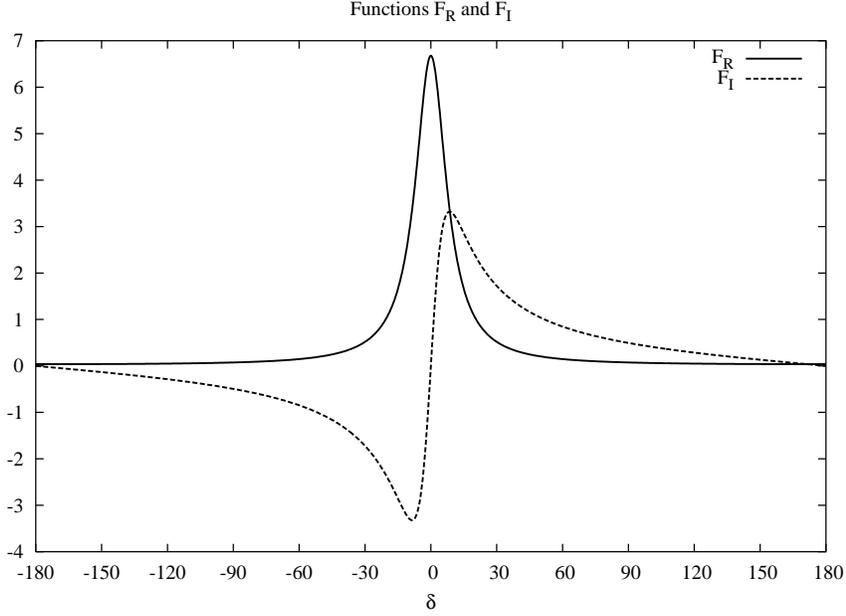}
%---------------------------------------
%\vspace*{-4mm}
\caption{The functions $F_R(\delta,d)$ and $F_I(\delta,d)$
versus the bunch to bunch phase $\delta$ for
$d=0.15$.}
\label{d015FRFI_plot}
\end{figure}

A plot of these functions is shown in Figure~\ref{d015FRFI_plot}
for a damping constant $d=0.15$.

Fundamentally, we are exciting a simple harmonic oscillator with a train of
$\delta$-function pulses, and $F_R(\delta,d)$ is proportional to the response
of the oscillator.  It shows a characteristic resonance bell-shape, with the
peak at the condition where the $\delta$ pulses arrive at any sub-harmonic of
the oscillator.  The functions $F_R(\delta,d)$ and $F_I(\delta,d)$ are 
asymptotic amplification factors along the bunch train; there is no 
bunch-to-bunch amplification of the energy deviation $\Delta E_n$ or kick
$\theta_n$ if these functions are smaller than one.

%-------------------------------------------------
%-------------------------------------------------

The average and the rms values of $F_R(\delta,d)$ and $F_I(\delta,d)$ for
$-\pi \leq \delta \leq \pi$ are:
\bea
 \left<F_R \right> & = & \frac{1}{2 \, \pi} \int_{-\pi}^{\pi} d\delta \, F_R(\delta,d) = \frac{1}{2} \\
		   \nonumber \\
 \left<F_I \right> & = & \frac{1}{2 \, \pi} \int_{-\pi}^{\pi} d\delta \, F_I(\delta,d) = 0 \\
		   \nonumber \\
\eea

By rewriting the finite series for the average of the absolute kick function
as a difference of two infinite series \footnote{We have 
$\sum_{k=1}^{\infty} \frac{1}{k} \exp(-k \, d) = - \ln(1-\exp(-d))$.}
we obtain
\bea
 \left<\Abs{F_I} \right>
                   & = & \frac{1}{\pi} \int_{0}^{\pi} d\delta \, F_I(\delta,d)
                     = \frac{1}{\pi} \ln(\coth(\frac{d}{2})) \\
		   & \approx & \frac{1}{\pi} \ln(\frac{2}{d}) \nonumber		   
\eea

Though the expectation $\left<F_I \right>$ in the interval $[-\pi,\pi]$ is zero,
the expectation for the absolute value of the kick gets larger as $d$ gets
smaller.
                      
Furthermore one obtains for the asymptotic RMS-values
\bea
{\rm RMS} (F_R) & = &
	\sqrt{\frac{1}{2\,\pi} \int_{-\pi}^{\pi} d\delta \, F_R(\delta,d)^2 }
		\nonumber \\
				& = &
	\frac{1}{2} \sqrt{\coth(d)}
		\nonumber \\
				& = &
	\sqrt{ \frac{1 + a^2}{4 (1-a^2)} } \\
{\rm RMS} (F_I) & = &
	\sqrt{\frac{1}{2\,\pi} \int_{-\pi}^{\pi} d\delta \, F_I(\delta,d)^2 }
		\nonumber \\
				& = &
	\frac{1}{2} \sqrt{(\coth(d) - 1)}
		\nonumber \\
				& = &
	\sqrt{ \frac{a^2}{2 (1-a^2)} }
		\nonumber \\
				& \approx &
	\frac{1}{2 \sqrt{d}},
\label{Eqn_FiRMS}
\eea
where the last approximation is valid in the small $d$ limit.

The RMS of $F_R$ around its mean of $1/2$ is equal to the RMS of $F_I$:
\bea
\sqrt{ {\rm RMS}(F_R)^2 - (1/2)^2 }  & = & {\rm RMS} (F_I) \nonumber \\
                & = & \frac{1}{2} \sqrt{ \coth(d) - 1} \\
                & \approx & \frac{1}{2 \, \sqrt{d}} - \frac{\sqrt{d}}{4} + \cdots
 \nonumber
\eea
The RMS of the function $F_R$ and $F_I$ grows as $1/\sqrt{d}$ and will
finally go to infinity if the damping is very small.

\begin{figure}[h!tbp]
\centering
\includegraphics*[width=120mm]{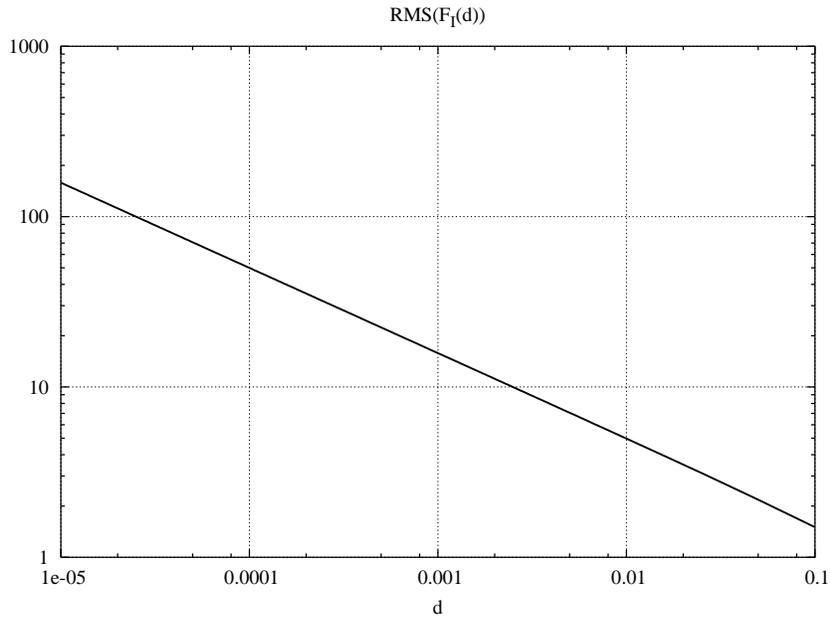}
%\vspace*{-4mm}
\caption{ The RMS of the asymptotic $F_I$ as a function of $d$.}
\label{hte_Frms_d}
\end{figure}

Figure \ref{hte_Frms_d} shows the dependence of $RMS(F_I)$
upon $d$.

In the no-damping limit,
\bea
  \lim_{d \rightarrow 0} F_R(\delta,d) 
			\rightarrow \frac{d}{2(1-d)} \, \frac{1}{1-\delta} \\ 
\nonumber 
\eea
which will go to $d/\delta^2$ for small $\delta$.  $F_I(\delta,d)$ approaches
the function
$f_I(\delta)$ which is
\bea
  f_I(\delta) & = & F_I(\delta,d=0) \nonumber \\
              & = & \frac{1}{2} \, \cot( \frac{\delta}{2} ) \\ 
\nonumber 
\eea
This form is the result of taking the limit of Equation \ref{Eqn_Sn}
first for large $n$ and then for small $d$.  Taking just the limit for small $d$,
\beq
S_{n}  =  \frac{ 1 - \exp(\imath (n-1)\delta) }{\exp(-\imath \delta) -1} 
\,\, {\rm for} \, \, d \rightarrow 0
\eeq
is showing that at very small dampings the deflection, i.e. the imaginary part of
$S_n$, will exhibit an oscillatory behavior before reaching equilibrium.

\begin{figure}[h!tbp]
\setlength{\unitlength}{1mm}
%---------------------------------------
\centering
%\dummyfig{./Eps/dvarFI.eps}
\includegraphics*[width=120mm]{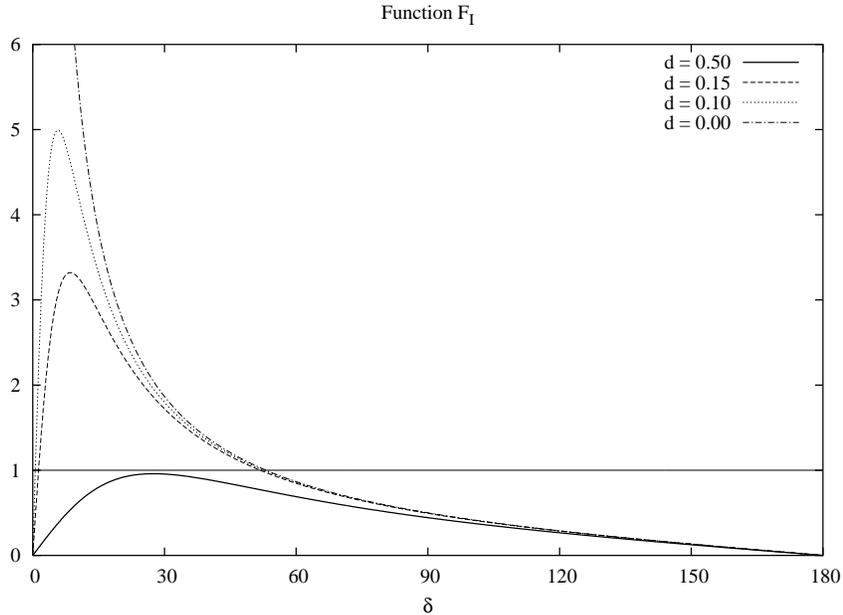}
%---------------------------------------
%\vspace*{-4mm}
\caption{The function $F_I(\delta,d)$ versus the bunch to bunch phase $\delta$
for different values of the damping constant $d$.}
\label{dvarFI_plot}
\end{figure}

Plots of function $F_I(\delta,d)$ for different values of the damping constant
$d$ ($0.1$, $0.15$, $0.5$) and the function $f_I(\delta) = F_I(\delta,d=0)$
are shown in Figure \ref{dvarFI_plot}.
%-------------------------------------------------
%------------------------------------------------- 

\begin{figure}[h!tbp]
\setlength{\unitlength}{1mm}
%---------------------------------------
\centering
%\dummyfig{./Eps/FImaxvd.eps}
\includegraphics*[width=120mm]{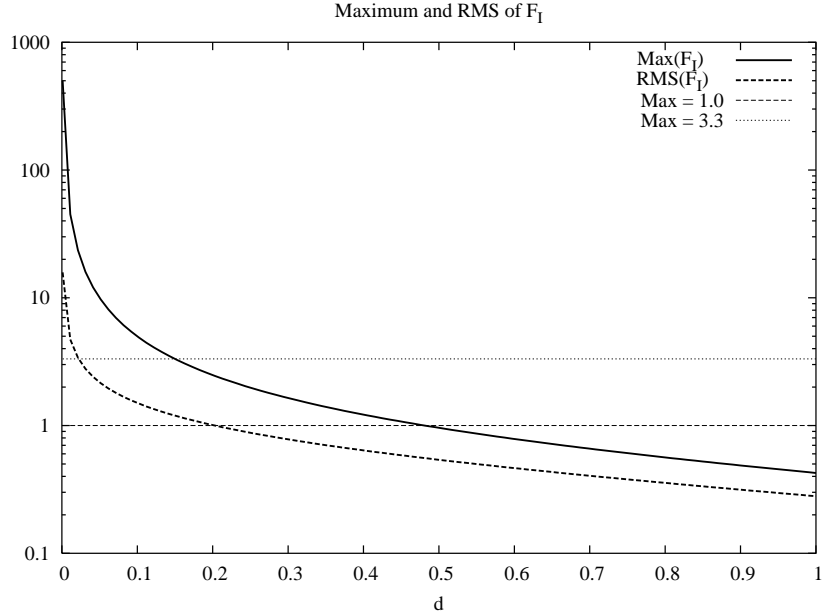}
%---------------------------------------
%\vspace*{-4mm}
\caption{The functions ${\rm Max}\left[F_I\right](d)$ and ${\rm RMS}(F_I)(d)$
 are plotted versus the damping constant $d$.}
\label{FImaxvd_plot}
\end{figure}

The maximum of the function $F_I$ is not at $\delta = 0$ like $F_R$ but at
\beq
  \delta_{\rm max} = \arccos(\frac{1}{\cosh(d)}) \approx d - \frac{d^3}{6} + \dots \hspace{3mm}.
\eeq
The maximum of the function at $\delta = \delta_{\rm max}$ is a function of the damping
constant $d$:
\beq
{\rm Max}\left[F_I\right](d) = 
F_I(\delta_{\rm max},d) = \frac{1}
                         {2 \, \sinh(d) }.
\eeq
Plots of functions ${\rm RMS}(F_I)(d)$ and ${\rm Max}\left[F_I\right](d)$
versus the damping constant $d$ are shown in Fig.~\ref{FImaxvd_plot}.
%-------------------------------------------------
%------------------------------------------------- 

The function $F_I(\delta,d)$ is smaller than one for all possible phases 
$\delta$ if the damping constant $d$ is equal or larger than
\beq
 d_1 = \ln\left( \frac{1}{2} \left( 1 + \sqrt{5} \right) \right) = 0.481 \hspace{2mm}.
\eeq
This corresponds to a Q-value of about $3 \times 10^4$ if a bunch to bunch 
spacing of $1/\Delta t = 1$~MHz is considered.  Such a low external Q-value is 
usually not achieved for all HOMs. But even if the damping constant is smaller
than $d_1$ the function $F_I(\delta,d)$ will be smaller than one for most 
phases (see Figure \ref{dvarFI_plot}).

For $d<d_1$ there exist two solutions of the equation
\beq
 F_I(\delta,d) = 1.
\eeq
The larger one may be denoted as $\delta_1$. We have
\beq
 \delta_1 \approx \arccos(3/5) - d^2 - \cdots \hspace{2mm}.
\eeq
The probability $P$ that $F_I(\delta,d)$ is larger than one is therefore
approximately
\beq
 P_{F_I>1} \approx \frac{1}{\pi} \, \arccos(3/5) \approx 0.3 .
\eeq
The amplification of the kick due to {\em one} HOM is expected to be smaller
than one in about 70~\% of the cavities of all possible random phases.
Since several modes have to be considered the 
overall result will be different from this optimistic one-mode
scenario (see the discussion in section \ref{sec-35}).

The probability that the function $F_I(\delta,d)$ is larger
than a constant $A$ can be estimated from the proability
that the function $f_I(\delta)$ is larger than $A$. In
general we have:
\beq
   P_{F_I>A} < P_{f_I>A} = \frac{2}{\pi} \,
               \arctan \left( \frac{1}{2 \, A} \right).
\eeq
For many practical cases it is $P_{F_I>A} \approx P_{f_I>A}$
if the constant $A$ is smaller than ${\rm Max}\left[F_I\right]$.

%--------------------------------------------------------------------
%--------------------------------------------------------------------
\subsection{The RMS kick over a bunch train due to one dipole mode}
\begin{figure}[h!tbp]
\setlength{\unitlength}{1mm}
%---------------------------------------
\centering
%\dummyfig{./Eps/Ave.eps}
\includegraphics*[width=120mm]{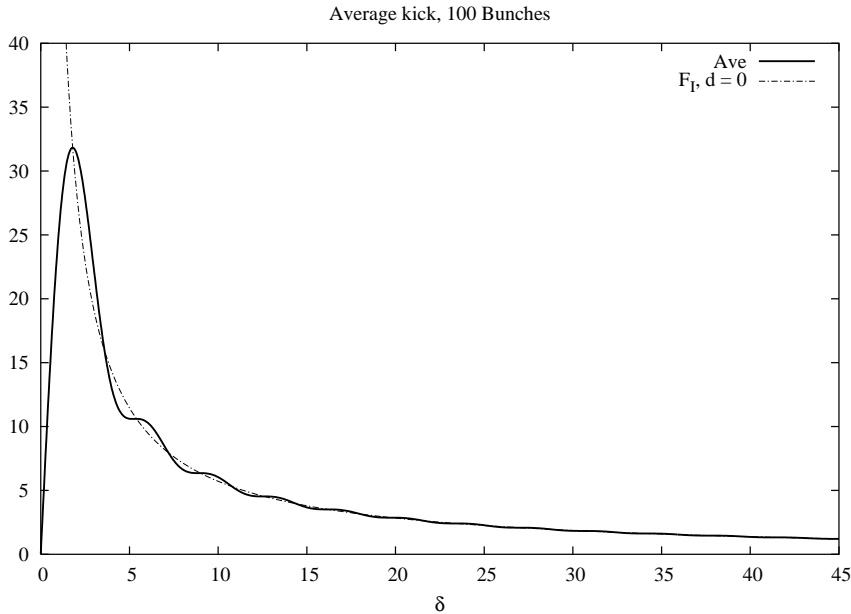}
%---------------------------------------
%\vspace*{-4mm}
\caption{The average kick function $Ave_b(N=100,\delta)$ and the asymptotic function 
$F_I(\delta,d = 0)$ is plotted versus the bunch to bunch phase $\delta$.}
\label{fvarAF}
\end{figure}

In the previous section we have discussed the asymptotic kick on the bunch 
train.  The average and RMS values of the asymptotic function $F_R(\delta,d)$ 
and $F_I(\delta,d)$ have been calculated for the phase $\delta$.  Now we drop
the asymptotic limit, and average the kick $\theta_n$ for finite bunch number 
$n$.  We give the RMS kick over the bunch train as well. We retain the full 
form of the kick
\beq
\theta_n / \hat{\theta}  =  F_I(\delta,d) - \Imag{R_n},
\eeq
where $F_I$ and $R_n$ are defined as in Equations \ref{FIdef} and \ref{Rndef}.
The average and the RMS of the kicks over $N$ bunches in the train, normalized
with respect to the kick amplitude $\hat{\theta}$ are
\bea
\left< \theta\right> / \hat{\theta} & = &  F_I - \frac{1}{N} \, \sum_{n=1}^{N} \, \Imag{R_n} \\
\nonumber \\
{\rm RMS}(\theta) / \hat{\theta} & = & \sqrt{ \frac{1}{N} \, \sum_{n=1}^{N} \, (\Imag{R_n})^2
                           - {\left( \frac{1}{N} \, \sum_{n=1}^{N} \, \Imag{R_n} \right)}^2}.
\eea
The analytic expression for the sums are:
\bea
\frac{1}{N} \, \sum_{n=1}^{N} \, \Imag{R_n} & = & 
\frac{{\rm e}^{- (N+2) \,d}}{2 N} \cdot
\frac{1}{ 1 +  {\rm e}^{-2 d} -2 \,{\rm e}^{-d} \,\cos(\delta) } \cdot \\
& & \frac{1}{\cosh(d)-\cos(\delta)} \cdot 
\left( {\rm e}^{N \, d} ({\rm e}^{2 \, d} -1) \sin(\delta) + \right.
\nonumber \\
\nonumber \\
& & \left. 2\, {\rm e}^{d} \sin(N \delta) 
- {\rm e}^{ 2d} \sin((N+1) \delta) -\sin((N-1) \delta) \right)
\nonumber \\
\nonumber \\
\frac{1}{N}\, \sum_{n=1}^{N} \, (\Imag{R_n})^2 & = &
\frac{ {\rm e}^{- (N+2) \,d} }{N} \cdot 
\frac{1}{ {(  1 +  {\rm e}^{-2 d} -2 \,{\rm e}^{-d} \,\cos(\delta)  )}^2 } \cdot \\
\nonumber \\
& & \left(-\cos(\delta) + \frac{\cosh(d)}{\sinh(d)} \, \sinh(N \,d) \right) + \nonumber \\
\nonumber \\
& & \frac{{\rm e}^{- 2(N+3) \,d}}{2 N} \cdot
\frac{1}{1 +  {\rm e}^{-2 d} -2 \,{\rm e}^{-d} \,\cos(\delta)^2} \cdot \nonumber \\
& &\frac{1}{1 +  {\rm e}^{-4 d} -2 \,{\rm e}^{-2 d} \,\cos(2 \,\delta)} \cdot \nonumber \\
\nonumber \\
& &\left( 2 \,{\rm e}^{(2 N + 1) \,d} ({\rm e}^{2\,d} -1 ) \cos(\delta)  + \right. \nonumber \\
& & {\rm e}^{2 N \,d} ({\rm e}^{4\,d} -1 ) \cos(2 \delta) - \cos(2\, (N-1) \,\delta) + \nonumber \\
& & {\rm e}^{4 \,d} \cos(2 \, (N+1) \, \delta) + 2 {\rm e}^{d} \cos((2N-1) \, \delta) -\nonumber \\
& & \left. 2 {\rm e}^{3 \,d} \cos((2N +1)\, \delta) \right) \nonumber
\eea
While the general analytic expression for the average and the rms kick over a
bunch train of $N$ bunches are rather complicated the expressions in the limit
of no damping at all ($d \rightarrow 0$) are much simpler:
\bea
Ave_b(N,\delta) & = & \lim_{d \rightarrow 0} \left< \theta\right> / \hat{\theta} \\
              & = & \frac{1}{2 \, (\cos(\delta) -1) }
                    \left( \frac{\sin(N \delta)}{N} - \sin{\delta} \right)
\nonumber \\
\nonumber \\
\nonumber \\
Rms_b(N,\delta) & = & \lim_{d \rightarrow 0} {\rm RMS}(\theta) / \hat{\theta} \\
& = & \frac{1}{2} \, \sqrt{
\frac{2\, (1-\cos(\delta)) - \frac{2 \, {\sin(N \delta)}^2}{N^2}  
             + \frac{\sin(2 \,N \, \delta) \, {\tan(\delta)}^2  }{N} }
        {{(1-\cos(\delta))}^2}
                          } \nonumber
\eea
Furthermore one may look for the limit of the functions
$Ave_b(N,\delta)$ and $Rms_b(N,\delta)$ for long bunch trains 
($N \rightarrow \infty$).  Provided the limit is taken with constant $\delta$,
so that$\delta N \rightarrow \infty$,
\bea
\lim_{N \rightarrow \infty} Ave_b(N,\delta) & = & \frac{1}{2} \cot( \frac{\delta}{2}) \\
      & = & \lim_{d \rightarrow 0} F_I(\delta,d) \nonumber \\
\nonumber \\
\lim_{N \rightarrow \infty} Rms_b(N,\delta) & = & \frac{1}{2\sqrt{2}} 
\frac{\sqrt{1-\cos(\delta)}}{{(\sin(\frac{\delta}{2}))}^2}.
\eea
The limit of $Ave_b(N,\delta)$ with respect to $N$ is equal to the asymptotic
amplification function  $F_I(\delta,d)$ in the limit of no damping, which 
diverges at $\delta =0$.  However, $Ave_b(N,\delta)$ is finite for all
$\delta \geq 1/N$ even if there is no damping.  The functions
$Ave_b(N,\delta)$ and $\lim_{d \rightarrow 0} F_I(\delta,d)$ are shown in 
Figure \ref{fvarAF} for a 100 bunch train.
%-------------------------------------------------
%-------------------------------------------------

%--------------------------------------------------------------------
%--------------------------------------------------------------------
\subsection{The RMS kick due to several dipole modes in an ensemble
of cavities}
\label{sec-35}
Based on the results of the previous subsections we want to calculate the expected
kick due several modes in an ensemble of cavities. We want to find damping
constants that limit the expected kick to less than the RMS beam divergence
at the cavity.  In this section, our estimates are based on the three dipole
modes of Table \ref{dipolmodes}. We assume that the electrical axis is
the same for all modes and defines the axis of the
cavity. The reference $r_0$ offset of the beam to the cavity
axis is always 1~mm. Furthermore we assume that the 
betatron-phase advance between the cavities is small and the kicks
from several cavities can be simply added.

\subsubsection{An estimate based on the asymptotic RMS kick}
\label{sec-351}
\begin{figure}[h!tbp]
\setlength{\unitlength}{1mm}
%---------------------------------------
\centering
%\dummyfig{./Eps/d0022FI.eps}
\includegraphics*[width=120mm]{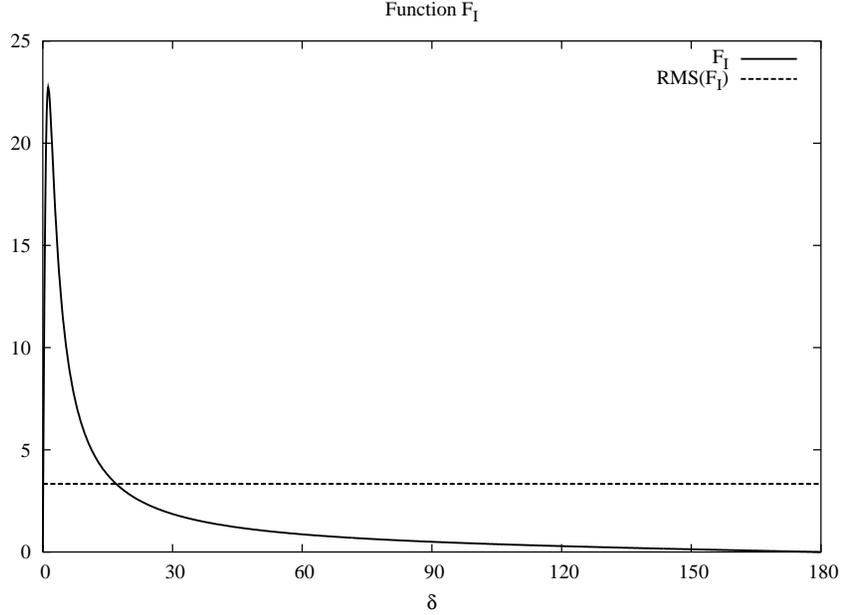}
%---------------------------------------
%\vspace*{-4mm}
\caption{The function $F_I(\delta,d)$
versus the bunch to bunch phase $\delta$ for
$d=0.0219$.  The RMS value ($3.33$) of the function
is also shown. The function $F_I(\delta,d)$ is larger than
the RMS value in the interval $[0.09^{\circ},16.9^{\circ}]$ }
\label{d0022FI_plot}
\end{figure}

An estimate using an RMS approach and including a sum square
combination of the different modes in the cavity is:
\beq
     {\theta_{RMS}}^2 = \sum_{k \in \{ {\rm modes} \}}
       {\hat{\theta}_k}^2 \, 
       \frac{1}{2 \, \pi} \int_{-\pi}^{\pi} d\delta \, {F_I(\delta,d_k)}^2.  
\eeq
If the damping constants of all modes are identical one obtains
for 4 cavities:
\bea
{\theta_{RMS}} & = & {\rm RMS} (F_I)(d) \, \cdot \,
                \sqrt{4 \, \sum_{k=1}^{3} {\hat{\theta}_k}^2} \\
\nonumber \\
               & = & {\rm RMS} (F_I)(d) \, \cdot \, 2.991 \, \mu{\rm rad} 
\label{rmstheta}
\eea
The value 2.991$\mu$rad is based on the values in table \ref{dipolmodes}.
Equation (\ref{rmstheta}) can be used as a criteria to determine
the damping constant $d$. If we demand that the combined RMS
kick should not exceed half of the single bunch divergence:
\beq
 2.991 \, \mu{\rm rad} \, \cdot \, {\rm RMS} (F_I)(d) < 10 \, \mu{\rm rad},
\label{onermsisten}
\eeq
one obtains for the required damping constant $d_{rms 10}$:
\beq
 d_{rms 10} = 0.0219\,\,. 
\eeq
The corresponding Q-values of the modes are listed in
Table \ref{Qvalued0022} and the function $F_I(\delta,d_{rms 10})$ is plotted in
Figure \ref{d0022FI_plot} for $0 \leq \delta \leq \pi$.
%%%%%%%%%%%%%%%%%%%%%%%%%%%%%%%%%%%%%%%%%%%%%%%%%%%%%%%%%%%%%%%%%%%%%%%%%%
\begin{table}[h!tb]
\centerline{
 \begin{tabular} {c||c||c}
$f$ / GHz  &  $Q_{ext}$            &  $Q_{ext}$            \\ 
           & ($1/\Delta t =1$~MHz) & ($1/\Delta t =5$~MHz) \\ 
 \hline
4.834 & $6.9 \times 10^5$ & $1.4 \times 10^5$ \\
5.443 & $7.7 \times 10^5$ & $1.5 \times 10^5$ \\
7.669 & $1.1 \times 10^6$ & $2.1 \times 10^5$ \\
\end{tabular}
}
\vspace{-0.1cm}
\caption{\label{Qvalued0022} 
Q-values of three modes for a damping constant $d=0.0219$
and bunch-to-bunch distances of $1000$~ns and $200$~ns.
}
\end{table}
%%%%%%%%%%%%%%%%%%%%%%%%%%%%%%%%%%%%%%%%%%%%%%%%%%%%%%%%%%%%%%%%%%%%%%%%%%
The RMS and maximum of the function $F_I(\delta,d)$ for the
damping constant
$d_{rms 10} = 0.0219$ are:
\beq
{\rm RMS} (F_I)(d_{rms 10}) = 3.33, \,\,\,\,\,\,
{\rm Max}\left[F_I\right](d_{rms 10}) = 22.9 \,\,.
\eeq
So in the worst but very unlikely case the total kick can be:
\bea
\theta_{tot} & = & {\rm Max}\left[F_I\right](d_{rms 10})  \cdot \,
                4 \, \sum_{k=1}^{3} {\hat{\theta}_k} \\
             & = &{\rm Max}\left[F_I\right](d_{rms 10})  \cdot \,
                 9.71 \,\mu{\rm rad} = 222.4 \,\mu{\rm rad}. \nonumber
\eea
The worst case assumes that the frequency of all 3 dipole modes in
all 4 cavities are tuned in a way that the bunch-to-bunch
phase is just equal to $\delta_{max}$ where the function
$F_I(\delta,d_{rms 10})$ has it maximum.
The probability that the function $F_I(\delta,d_{rms 10})$
is larger than the RMS value is: 
\beq
 P_{F_I>3.33} = 9.35 \, \%.
\eeq
%-------------------------------------------------
%-------------------------------------------------

%%%%%%%%%%%%%%%%%%%%%%%%%%%%%%%%%%%%%%%%%%%%%%%%%%%%%%%%%%
\subsubsection{Alternative constraints on the asymptotic kick}
\label{sec-352}
Instead of constraining the RMS kick one may put a constrain
on the RMS value of the maximum possible kicks:
\beq
     {\Theta_{RMS}}^2 = \sum_{k \in \{ {\rm modes} \}}
       {\hat{\theta}_k}^2 \, \,{\left({\rm Max}\left[F_I\right](d_k)\right)}^2.  
\eeq
If the damping constants of all modes are identical one obtains
for 4 cavities:
\bea
{\Theta_{RMS}} & = & {\rm Max}\left[F_I\right](d) \, \cdot \,
                \sqrt{4 \, \sum_{k=1}^{3} {\hat{\theta}_k}^2} \\
\nonumber \\
           & = & {\rm Max}\left[F_I\right](d) \, \cdot \, 2.991 \, \mu{\rm rad} 
\nonumber \\
           & < & 10 \mu{\rm rad}
\label{rmsTheta}
\eea
The RMS value of the maximum possible kicks is smaller than $10 \mu{\rm rad}$
if $d = d_{RMS} = 0.15$, since 
\beq
{\rm Max}\left[F_I\right](d_{RMS}) < 3.34 \,\,.
\eeq
The corresponding Q-values are summarized in Table \ref{Qvalued015}.
%%%%%%%%%%%%%%%%%%%%%%%%%%%%%%%%%%%%%%%%%%%%%%%%%%%%%%%%%%%%%%%%%%%%%%%%%%
\begin{table}[h!tb]
\centerline{
 \begin{tabular} {c||c||c}
$f$ / GHz  &  $Q_{ext}$            &  $Q_{ext}$            \\ 
           & ($1/\Delta t =1$~MHz) & ($1/\Delta t =5$~MHz) \\ 
 \hline
4.834 & $1.0  \times 10^5$ & $2.0 \times 10^4$ \\
5.443 & $1.14 \times 10^5$ & $2.3 \times 10^4$ \\
7.669 & $1.6  \times 10^5$ & $3.2 \times 10^4$ \\
\end{tabular}
}
\vspace{-0.1cm}
\caption{\label{Qvalued015} 
Q-values of three modes for a damping constant $d=0.15$
and bunch-to-bunch distances of $1000$~ns and $200$~ns.
}
\end{table}
%%%%%%%%%%%%%%%%%%%%%%%%%%%%%%%%%%%%%%%%%%%%%%%%%%%%%%%%%%%%%%%%%%%%%%%%%%
The RMS of the function $F_I(\delta,d)$ is $1.2$ and therefore
\bea
{\theta_{RMS}} & = & {\rm RMS} (F_I)(d_{RMS}) \, \cdot 
                    \, 2.991 \, \mu{\rm rad} \nonumber \\
               & = & 3.59 \, \mu{\rm rad} \,\, , \\
 \nonumber \\
\theta_{tot} & = &{\rm Max}\left[F_I\right](d_{RMS})  \cdot \,
                 9.71 \,\mu{\rm rad} \nonumber \\
             & = & 32.2 \,\mu{\rm rad} \,\,.
\eea
Even in the worst case the total kick will only be a factor $1.5$
larger than the single beam divergence of $20 \,\mu{\rm rad}$.

%%%%%%%%%%%%%%%%%%%%%%%%%%%%%%%%%%%%%%%%%%%%%%%%%%%%%
One of the most strict alternatives is to constrain the
worst case of the total kick to $10 \, \mu{\rm rad}$:
\bea
\theta_{tot} & \leq & {\rm Max}\left[F_I\right](d) \, \cdot \,
                      4 \, \sum_{k=1}^{3} \hat{\theta}_k \\
             & = & {\rm Max}\left[F_I\right](d) \, \cdot \, 
                     9.71 \, \mu{\rm rad} \nonumber \\
             & \leq & 10 \, \mu{\rm rad} \nonumber
\eea
If the damping constant $d$ is larger $d_{max} = 0.47$
we obtain
\beq
{\rm Max}\left[F_I\right](d) \leq 1.03 \,\,,
\eeq
and the constrain $\theta_{tot} \leq  10 \, \mu{\rm rad}$
is met. The corresponding Q-values are listed in Tab.~\ref{Qvalued047}.
For this very strict constraint we have:
\beq
{\theta_{RMS}}  =  {\rm RMS} (F_I)(d_{max}) \, \cdot 
                    \, 2.991 \, \mu{\rm rad}
                =  1.69 \, \mu{\rm rad} \,\, .
\eeq
%%%%%%%%%%%%%%%%%%%%%%%%%%%%%%%%%%%%%%%%%%%%%%%%%%%%%%%%%%%%%%%%%%%%%%%%%%
\begin{table}[h!tb]
\centerline{
 \begin{tabular} {c||c||c}
$f$ / GHz  &  $Q_{ext}$            &  $Q_{ext}$            \\ 
           & ($1/\Delta t =1$~MHz) & ($1/\Delta t =5$~MHz) \\ 
 \hline
4.834 & $3.2 \times 10^4$ & $6.5 \times 10^3$ \\
5.443 & $3.6 \times 10^4$ & $7.3 \times 10^3$ \\
7.669 & $5.1 \times 10^4$ & $1.0 \times 10^4$ \\
\end{tabular}
}
\vspace{-0.1cm}
\caption{\label{Qvalued047} 
Q-values of three modes for a damping constant $d=0.47$
and bunch-to-bunch distances of $1000$~ns and $200$~ns.
}
\end{table}
%%%%%%%%%%%%%%%%%%%%%%%%%%%%%%%%%%%%%%%%%%%%%%%%%%%%%%%%%%%%%%%%%%%%%%%%%%

%% file: toymc.tex
% ############################################################
%
% FILE: toymc.tex
%
% ############################################################
%
\section{Monte Carlo Analysis}
\label{sec-4}

The problem has also been attacked computationally with a simple
Monte Carlo calculation.  A worst-case analysis for a single cavity, 
and typical-case analyses for both FLASH and XFEL configurations have
been made.  The analysis provides, with certain assumptions, $Q_{ext}$ 
requirements.

The analysis does allow for the scatter of HOM frequencies that will
no doubt result from variations within tolerances of the cavity 
dimensions and shapes that are part of the manufacturing tolerances.
The analysis does not allow for the possibility that if HOM dampers
fail utterly, fields may remain in the cavity for the relatively long
time between bunch trains.
Perhaps most importantly, there will be some active-feedback beam 
steering system that should ameliorate beam deflection effects and this
remains still to be studied.

%----------------------------------------------------------------------------
%----------------------------------------------------------------------------
\subsection{Method}

The analysis is best explained by describing the objects from which it is made.
A {\tt mode} is basically a frequency, an $\frac{ R^{(m)}}{Q}$ value, an 
azimuthal quantum number and a decay time.  A {\tt cavity} is a collection of
modes and the corresponding wakefield functions $W_{\|}$ and
$W_{\bot}$.  A {\tt beamline} is a set of cavities, with the kicks and energy
losses from each of the cavities summed.

A technical note: the algorithm directly implements Equation \ref{wake1} and for
the bunch train lengths under consideration here, that requires taking the 
$\sin(\theta)$ and $\cos(\theta)$ for very large values of $\theta$ indeed. 
Care has been taken to verify the numeric precision of these functions in this
specific implementation of the algorithm.

All the modes listed in Table~\ref{dipolmodes} are included in the simulation,
although the quadrupole mode has a negligible effect.  The effect of the 
monopole mode on the beam energy is computed, although we have concentrated on
the transverse dynamics.  A sixth entry exists in the code for the 3.9~GHz
operating mode, but it is switched off and is not used.

%----------------------------------------------------------------------------
%----------------------------------------------------------------------------
\subsection{Single cavity worst-case analysis}

The worst-case is where a cavity has a HOM with a peak in $F_I$
that is directly on beam resonance and the external dampers have failed 
so that damping is provided by the intrinsic power dissipation of the Nb
surface alone.  In this case, the asymptotic limit
is not reached; the bunch train is not long relative to the wakefield time
scale.  The surface resistance is modeled from $3.9$~GHz with an $R_{BCS}$ of 
$20n\Omega$ scaled with the square of the frequency ratio, plus the residual
resistance, $R_0$, of $20n\Omega$.  The $Q$-value is then obtained by dividing 
$G_1$ of the mode by this resistance.

\begin{figure}[h!tbp]
\centering
\includegraphics*[width=120mm]{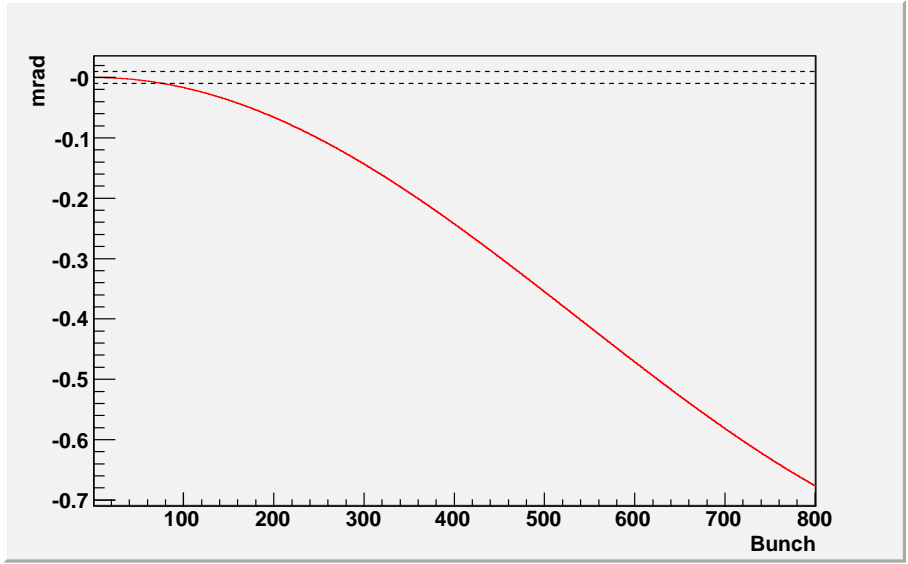}
%\vspace*{-4mm}
\caption{Bunch deflection as a function of bunch number, worst case
analysis of the $4.834$~GHz mode.  For this mode, $Q = 5.4 \times 10^9$, 
$d = 2.77 \times 10^{-6}$, and $\hat{\theta} = 1.22$~mrad.}
\label{worst_deflplot}
\end{figure}

\begin{figure}[h!tbp]
\centering
\includegraphics*[width=120mm]{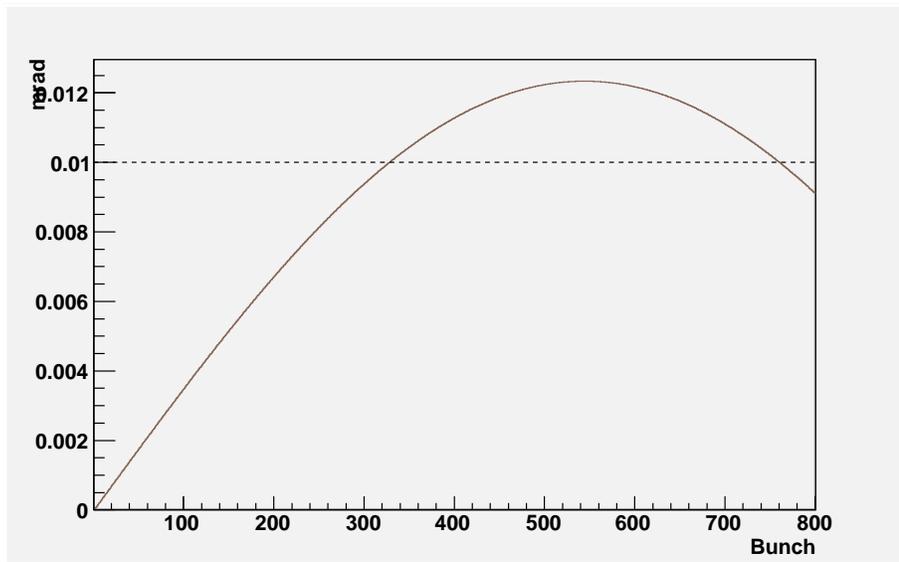}
%\vspace*{-4mm}
\caption{Bunch crabbing as a function of bunch number, worst case
analysis of the $4.834$~GHz mode.}
\label{worst_crabplot}
\end{figure}

The monopole mode will, with over 800 bunches in the FLASH configuration, take 
about $800\kev$ out of the last bunch in the train.  The maximum of $F_I$ 
corresponds to about $468\hz$ from the exact on-resonance condition for all
of the dipole modes~\footnote{In the long bunch train limit, the maximum would
occur at $\Delta f = f/{2Q}$.  However, without damping, 800 bunches is not near
the long bunch train limit.  Additionally, there are large-N oscillation effects,
as described in section \ref{sec-32}.}.  The deflection from the dipole mode of
highest $\frac{ R^{(m)}}{Q}$ in this case is shown in Figure \ref{worst_deflplot}; the
other modes have the same shape but with the vertical scale proportional to
$\frac{ R^{(m)}}{Q}$.  Figure~\ref{worst_crabplot} shows the 
{\it crabbing angle}, defined as the deflection angle (in the lab frame) at
the $1\sigma_z$ bunch head minus the deflecting angle at the bunch center.  This
is computed by changing, in effect, the trailing distance of the witness by
one bunch-length, and taking the  difference in displacement from the centroid
displacement.

We do not have a clear understanding as to how this can effect the lasing
process, but note that the angles involved are typically smaller than the
deflecting angles.  The energy loss in this dipole mode at this 
pseudo-resonance condition is only $5keV$.

It is clear that even a single cavity hitting the worst-case will cause 
lasing to stop.

%----------------------------------------------------------------------------
%----------------------------------------------------------------------------
\subsection{Typical-case analysis, FLASH beamline}

In the typical-case analysis we use the Monte Carlo method to examine the
probability distribution for beamline performance. 

A virtual beamline is constructed of 4 cavities.  The frequencies of the modes
are shifted from the nominal simulation results of Table~\ref{Tbl_modes} by
drawing on a uniform random distribution.  The width of the distribution 
spans the full range of $\delta$, corresponding to a frequency shift of
$-0.5$~MHz to $+0.5$~MHz in the FLASH case, and 1/5th that for the XFEL
injector.  While we do not have enough cavities to really check, this is
thought to be about the scale on which the scatter will actually be.

Simulated results such as in Figures \ref{worst_deflplot} and 
\ref{worst_crabplot}, along with the corresponding energy loss plot are
recorded.
%; an example of such a result, showing clear interference effects 
%between a pair of near-resonance modes is given in Figure \ref{perHTE}.
Then a new beamline, constructed with new calls to the pseudo-random number
generator is created and again, variation of deflection, crabbing, and energy loss
with respect to bunch number is determined.  The process is repeated 5000 
or in some cases, 10000 times.  At each bunch number, we have an average and an 
RMS, over those beamlines, of deflection, crabbing, and energy loss.

% Leo, a note to yourself:
% The Xcode projects used were ToyMC, BBL and XFEL, in that order.

%\begin{figure}[h!tbp]
%\centering
%\includegraphics*[width=13cm]{./Eps/example.eps}
%%\vspace*{-4mm}
%\caption{Deflection profile as a function of bunch number, for a virtual FLASH 
%beamline with a randomly selected frequency for the first dipole
%mode and no HOM damping.}
%\label{perHTE}
%\end{figure}

The average deflection over an ensemble is easily seen to be zero, as deflection
to the left is just as likely as deflection to the right.  The average energy
loss is also zero; the $\sin$ of Equation \ref{wake1}b is replaced by the $\cos$
of Equation \ref{wake1}a.
The widths of the distributions of deflection angle, crabbing angle and energy
loss describe at the $1\sigma$ level what kind of performance one can expect
when the machine is actually turned on.

Clearly, a design that will keep deflection down to our $\pm10\mu$rad goal at 
the 68.27\% (1$\sigma$) confidence level is risky; we need the RMS to be well
below the goal.  How far below is a difficult decision involving overall
program risk.  Notwithstanding, a statistical analysis is able to provide us
with a good sense of what levels of external damping we need to have.

\begin{figure}[h!tbp]
\centering
\includegraphics*[width=120mm]{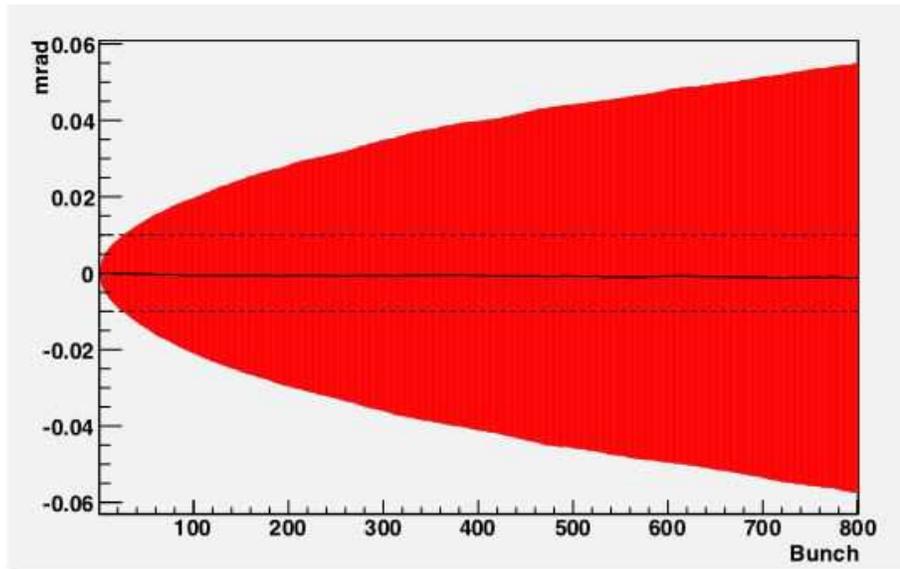}
%\vspace*{-4mm}
\caption{ The one-sigma deflection profile as a function of bunch number, averaged
over ensemble of 5000 virtual FLASH beamlines without external dampers.}
\label{stats_bad}
\end{figure}

Again, we do not obtain an acceptable result on just cavity self-damping alone.
If the HOM design fails across the board, deflections of $\sim\pm50\mu$rad
appear at the $1\sigma$ level by the end of an 800 bunch train in the 4 cavity,
5 mode FLASH model, as shown in Figure ~\ref{stats_bad}.  Next, we determine
how far we smust lower $Q_{ext}$ to stabilize the beam.

% Note from Helen:
%The multiplier of Fi from the Monte Carlo is multiplied by SQRT(4*Sum Theta-hat^2)
%= 2.99 as in Eqns 63, 64. These results can be compared with what one would
% obtain from Fi(RMS) in Figure(my/Leo fig). Points on Fig 12 indicate this 
%comparison. 
%Bunches      microrad
%10           6
%100          20
%200          30
%300          36
%500          42
%800          60
%the above are just eye ball. I believe a fit is something like Fi RMS=0.71SQRT(n-1). n-1=1, F=0.71; n-1=9, F=2.12; n-1=99, F=7.03; n-1=999, F=22.2
% Maybe 0.71 is 1/sqrt2?

For the $7.506\mhz$ monopole mode, lowering $Q_{ext}$ to a bit below 
$1 \times 10^7$ causes the RMS of $\Delta E$ to flatten out at large bunch
numbers to about $23\kev$.

For the dipole modes, there is some allocation of the $\pm10\mu$rad
target into the three modes; we have to also allocate fractions of the
deflection budget to the different deflecting modes.  If we require that the
RMS deflection flattens out to a level where $\pm10\mu$rad corresponds to
$3\sigma$ of the total deflection, and then allocate the deflection budget
evenly among our three large dipole modes, then each mode must contribute
$10\mu$rad$ / (3 \sqrt(3)) = 1.9\mu$rad in the asymptotic limit for the 4
cavities, as the contributions from the included modes are added in quadrature.

The Monte Carlo calculation differs from the analytic form of 
section \ref{sec-351} in that
\begin{enumerate}
\item  In the MC method, the asymptotic limit is not assumed.  It will in most
		cases be justified because the damping is strong.
\item  In the MC method, angles are summed over modes and cavities, and then an
		RMS is taken; in the analytic form, the sequence of these two
		operations is inverted.
\item  In the analytic form, the damping constants of all the modes are
		forced to be equal; in the MC method, the deflection in the
		long-bunchtrain limit are taken to be equal.
\item  Equation \ref{onermsisten} sets $10 \mu{\rm rad}$ to 1 times the RMS
		deflection; section \ref{sec-352} discusses alternate, tighter
		constraints.  The MC method uses a $3 \sigma$ constraint.
\end{enumerate}

The Monte Carlo results can be reproduced with analytically.  With the required
values of $Q_{ext}$, the asymptotic form of Equation \ref{Eqn_FiRMS} permits
direct solution for $Q_{ext}$ using the definition of $\hat{\theta}$ and $d$.
These relations lead to $Q_{ext}$ for the 3 modes of $3.7 \times 10^4$,
$2.5 \times 10^5$ and $1.7 \times 10^5$, or about 70-80\% of the MC values.

\begin{figure}[h!tbp]
\centering
\includegraphics*[width=120mm]{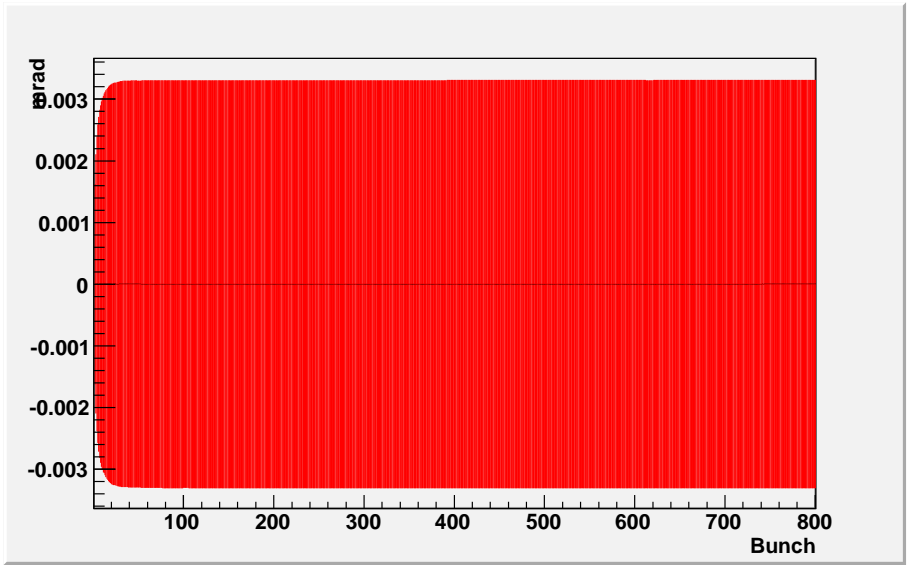}
%\vspace*{-4mm}
\caption{The one sigma deflection profile as a function of bunch number, averaged over 
ensemble of 10000 virtual FLASH beamlines, each consisting of 4 cavities and 5
well-damped modes.}
\label{FLASHouldwork}
\end{figure}

The $Q_{ext}$ values that produce deflections of the specified level in the
Monte Carlo, found basically by trial and error, are shown in Table \ref{qextmc},
along with the results of section \ref{sec-351} when a $3 \sigma$ constraint is used,
the results of Phillipe Piot's (unpublished) calculation and the recent 
measurements from prototypes \cite{TimerMeasures}.
The system is well damped in the simulation, as shown in Figure
\ref{FLASHouldwork}.

\begin{table}[ht]
\begin{center}
\begin{tabular}{cc|cccc}
Frequency & Azimuthal & \multicolumn{4}{c} {$Q_{ext}$}                       \\
   (GHz)  &  number   &
                  MC method    & $3\sigma$ Analytic &     Piot spec.   &     Measured   \\
\hline
 7.506  & 0 & $1.0\times 10^7$ &                    &                  & $1\times 10^4$ \\
 4.834  & 1 & $5.4\times 10^4$ & $9.0\times 10^4$   & $2.8\times 10^3$ & $1\times 10^4$ \\
 5.443  & 1 & $2.9\times 10^5$ & $1.0\times 10^5$   & $1.0\times 10^5$ & $5\times 10^4$ \\
 7.668  & 1 & $2.1\times 10^5$ & $1.4\times 10^5$   & $2.5\times 10^5$ & $1\times 10^4$ \\
\hline
\end{tabular}
\vspace{-0.1cm}
\caption{\label{qextmc}
$Q_{ext}$ requirements from this Monte Carlo study, the analytic forms 
of \ref{sec-351} when calculated for a $3 \sigma$ constraint, Phillipe Piot's 
analysis, and measurement from prototypes.  The FLASH beam parameters are used.
The value of $d_{rms10}/3$ is 0.169.}
\end{center}
\end{table}

%----------------------------------------------------------------------------
%----------------------------------------------------------------------------
\subsection{Typical-case analysis, XFEL injector beamline}

The typical-case analysis for our canonical XFEL configuration shows, as
expected, that the RMS deflection due to HOMs varies inversely as the beam
energy and approximately as the square root of the number of cavities.  This
makes the deflection about 2/3 of what it would be in FLASH at the same
bunch spacing.

Moving the bunch frequency up to $5\mhz$ changes the relative contribution of
the dipole modes by changing $d$ and thereby increases the accumulated
deflections.  Where the asymptotic RMS values of the deflections, scaled to
24 cavities in a $500\mev$ beam would otherwise be $1.2\mu$rad, the 
introduction of $5\mhz$  bunch frequency results in deflections of about 
$2.8\mu$rad.  To restore $3 \sigma$~$10\mu$rad performance, we need the
$Q_{ext}$ values of table \ref{XFELqext}.

\begin{table}[ht]
\begin{center}
\begin{tabular}{cc|cc}
Frequency & Azimuthal & \multicolumn{2}{c} {$Q_{ext}$}                       \\
   (GHz)  &  number   &
                  MC method    & $3\sigma$ Analytic \\
\hline
 7.506  & 0 & $1.0\times 10^7$ &                    \\
 4.834  & 1 & $2.4\times 10^4$ & $1.8\times 10^4$   \\
 5.443  & 1 & $1.3\times 10^5$ & $2.0\times 10^4$   \\
 7.668  & 1 & $1.0\times 10^5$ & $2.8\times 10^4$   \\
\hline
\end{tabular}
\vspace{-0.1cm}
\caption{\label{XFELqext}
$Q_{ext}$ requirements from this Monte Carlo study and the analytic forms 
of \ref{sec-351} when calculated for a $3 \sigma$ constraint.  The XFEL beam
parameters are used.}
\end{center}
\end{table}

The overall deflection profile still shows a well-damped system; the asymptotic
one sigma deflection of $3.39\mu$rad is reached within 60 bunches.  The asymptotic
energy change is $\sim 58\kev$.

%% file: conclusion.tex
% ############################################################
%
% FILE: conclusion.tex
%
% ############################################################
%
\section{Conclusion}
\label{sec-5}

We have studied the $Q_{ext}$ requirements for the ``third harmonic" cavities
using sets of beam parameters typical of FLASH and XFEL operation.  A key 
assumption is that lasing will cease when the deflection due to wakefields
approaches the divergence of the beam at the cavity location.  We have taken 
the divergence to be $\pm 10\mu$rad for both beam parameter sets.  The XFEL
optics in the vicinity of the ``third harmonic" cavities is not finalized
as we write, but the $\pm 10\mu$rad condition is of the correct scale.

The reader need also be aware of the 'program risk' issue: what level of
statistical confidence that this $\pm 10\mu$rad specification be met without
component replacement or change of beam parameters is required?  Here, we have
selected a $3\sigma$ level of confidence.  One might choose to argue that the
$3\sigma$ choice is too conservative.  Choosing a $1\sigma$ requirement relaxes
the damping requirements by about an order of magnitude.

The results are based basically on three modes of high beam-cavity coupling.
Adding in quadrature a number of other modes with lower $R^{(1)}/Q$ has little
influence ($\sim$25\%) on the damping required.

Both analytic and Monte-Carlo based analysis have been done.  Both allow for 
manufacturing defects, but do not allow for the action of any kind of active 
beam steering system and are quite conservative in that regard.

The results of the two analyses are broadly consistent and are summarized in
tables \ref{qextmc} and \ref{XFELqext}.  It is encouraging that the $Q_{ext}$
values measured on prototypes are better than the required values.

For damping values of $Q_{EXT}$ in the $10^5$ range, which are typical for 
functioning HOM mode dampers, the deflections reach their asymptotic values
after a few 10's of bunches.  In this case, our analytic results are
particularly easy to use, and we summarize them here.

The angular kick on bunch $n \in \{1,2,3\ldots\}$ due to a single mode of
angular frequency $\omega$ and quality number $Q$ in a train with bunches
$\Delta t$ seconds apart is $\theta_n = \hat{\theta} F_I(\delta,d)$ where 
\bea
\hat{\theta}_n & = &
	\frac{e \, q_{bunch}}{E_{beam}} \, c \, \frac{R^{(1)}}{Q} \, r_0 \nonumber\\
\delta & = & \omega \, \Delta t                                      \nonumber\\
d & = & \frac{\omega}{2 \, Q} \, \Delta t.                           \nonumber
\eea

The function $F_I$, which describes the bunch-to-bunch amplification of the
wakes, has a maximum at
\bea
\delta_{\rm max} & = &
	\arccos(\frac{1}{\cosh(d)}) \approx d - \frac{d^3}{6} + \dots    \nonumber\\
F_I(\delta_{\rm max},d) & = & \frac{1}{2 \, \sinh(d) }               \nonumber
\eea
and an RMS of
\bea
{\rm RMS} (F_I) & = &
	\sqrt{\frac{1}{2\pi} \int_{-\pi}^{\pi} d\delta F_I(\delta,d)^2 } \nonumber\\
				& = &
	\frac{1}{2} \sqrt{(\coth(d) - 1)}.                               \nonumber
\eea

\section{Acknowledgement}

We thank Don Edwards for an illuminating conversation.

%% file: references.tex
% ###########################################################
%
% FILE: references.tex
%
% ###########################################################
\addcontentsline{toc}{section}{\protect\numberline{}{References}}
%

%% file: master.bbl
\begin{thebibliography}{99.}
%
\bibitem{Wan01}
R.~Wanzenberg,
{\it Monopole, Dipole and Quadrupole Passbands of the TESLA 9-cell cavity},
TESLA 2001-33, Sept. 2001
%
\bibitem{Wakes}
T.~Weiland, R.~Wanzenberg,
{\it Wake fields and impedances},
in: Joint US-CERN part. acc. school, Hilton Head Island, SC, USA,
7 - 14 Nov 1990 / Ed. by M Dienes, M Month and S Turner. - Springer,
Berlin, 1992- (Lecture notes in physics ; 400) - pp.39-79 
%
\bibitem{MAFIA}
T.~Weiland,
{\it On the numerical solution of Maxwell's Equations and
Applications in the Field of Accelerator Physics},
Part. Acc. 15 (1984), 245-292
%
\bibitem{CST}
{\it MAFIA Release 4 (V4.021) }
CST GmbH, B\"udinger Str. 2a, 64289 Darmstadt, Germany
%
\bibitem{Wei83}
T.~Weiland,
{\it Comment on wake field computation in time domain},
DESY M-83-02, Feb. 1983
%
\bibitem{wilson82}
P.B.~Wilson
{\it High Energy Electron Linacs:
Application to Storage Ring RF Systems and Linear Colliders},
AIP Conference Proceedings 87, American Institute of Physics,
New York (1982),p. 450-563
%
\bibitem{PaWe56}
W.K.H.~Panofsky, W.A.~Wenzel,
{\it Some considerations concerning the transverse deflection
of charged particles in radio-frequency fields },
Rev. Sci. Inst. Vol 27, 11 (1956), 967
%
\bibitem{TimerCalculates}
W.F.O.~M\"{u}ller, J.~Sekutowicz, R.~Wanzenberg and T.~Weiland,
{\it A Design of a 3rd Harmonic Cavity for the TTF 2 Photoinjector},
TESLA-FEL 2002-05, July 2002.
T.~Khabiboulline, N.~Solyak and R.~Wanzenberg,
{\it Higher order modes of a 3rd harmonic cavity with an
increased end-cup iris}
FERMILAB-TM-2210, DESY-TESLA-FEL-2003-01, May 2003. 
%
\bibitem{Abramowitz}
M.~Abramowitz, I.A.~Stegun, (eds.)
{\it Handbook of Mathematical Functions}, 9th printing,
Dover, New York 1970
%
\bibitem{TimerMeasures}
T.~Khabiboulline, private communication.

%-------- These are not actually used -----------------
%%
%\bibitem{Aune:2000gb}
%B.~Aune {\it et al.},
%{\it The superconducting TESLA cavities},
%Phys.\ Rev.\ ST Accel.\ Beams{\bf 3} (2000) 092001
%[physics/0003011].
%%%CITATION = PHYSICS 0003011;%%
%%
%\bibitem{URMEL}
%T. Weiland,
%{\it On the computation of resonant modes in cylindrically
% symmetric cavities},
%NIM 216 (1983) 329-348
%%
%\bibitem{Watkins58}
%D.A.~Watkins
%{\it Topics in Electromagnetic Theory },
%Wiley 1958, New York
%%
%\bibitem{Handbook79}
%.R.~Lide, Ed.
%{\it Handbook of Chemistry and Physics },
%79th edition, 1998-1999, CRC Press, Washington, D.C.
%%
%\bibitem{PaPh62}
%W.K.H.~Panofsky, M.~Phillips,
%{\it Classical Electricity and Magnetism},
%Addison-Wesley, Reading Massachusetts 1962
%%
%\bibitem{Padamsee98}
%H.~Padamsee, J.~Knobloch, T.~Hays,
%{\it RF Superconductivity for Accelerators},
%Wiley, New York 1998

\end{thebibliography}
